\documentclass[10pt]{article}

\usepackage{url}            
\usepackage{booktabs}       
\usepackage{amsfonts}       
\usepackage{nicefrac}       
\usepackage{microtype}      

\usepackage{amsbsy}
\usepackage{amsmath}
\usepackage{amsthm}

\usepackage{color}
\usepackage[all]{xy}

\usepackage{wrapfig}

\usepackage{epsf}
\usepackage{epsfig}

\usepackage[switch]{lineno}  %

\usepackage{enumitem}

\usepackage[vlined,linesnumbered,ruled,resetcount]{algorithm2e}

\usepackage{bbm}

\usepackage[margin=1.2in]{geometry}

\SetKwInput{KwInput}{Input}                
\SetKwInput{KwOutput}{Output}              

\SetCommentSty{mycommfont}

\newcommand{\bfmu}{\boldsymbol{\mu}}
\newcommand{\bfzero}{\boldsymbol{0}}
\newcommand{\bfI}{\boldsymbol{I}}

\newcommand{\bfW}{\boldsymbol{W}}
\newcommand{\bfX}{\boldsymbol{X}}
\newcommand{\bfx}{\boldsymbol{x}}
\newcommand{\bfY}{\boldsymbol{Y}}
\newcommand{\bfy}{\boldsymbol{y}}
\newcommand{\bfc}{\boldsymbol{c}}
\newcommand{\bfw}{\boldsymbol{w}}

\newcommand{\bfD}{\boldsymbol{D}}
\newcommand{\bfK}{\boldsymbol{K}}

\newcommand{\bfi}{\boldsymbol{i}}
\newcommand{\bfj}{\boldsymbol{j}}
\newcommand{\bfai}{\boldsymbol{i}^\ast}
\newcommand{\bfaj}{\boldsymbol{j}^\ast}

\newcommand{\bfaX}{\boldsymbol{X}^\ast}
\newcommand{\bfaY}{\boldsymbol{Y}^\ast}

\newcommand{\bfax}{\boldsymbol{x}^\ast}
\newcommand{\bfay}{\boldsymbol{y}^\ast}

\newcommand{\ind}{\mathbbm{1}}
\newcommand{\bfind}{\boldsymbol{1}}

\title{Computationally efficient permutation tests for the multivariate two-sample problem based on energy distance or maximum mean discrepancy statistics}

\author{Elias Chaibub Neto}

\date{Sage Bionetworks, Seattle, WA 98121, USA \\ elias.chaibub.neto@sagebase.org}

\begin{document}

\maketitle

\begin{abstract}
Non-parametric two-sample tests based on energy distance or maximum mean discrepancy are widely used statistical tests for comparing multivariate data from two populations. While these tests enjoy desirable statistical properties, their test statistics can be expensive to compute as they require the computation of 3 distinct Euclidean distance (or kernel) matrices between samples, where the time complexity of each of these computations (namely, $O(n_{x}^2 p)$, $O(n_{y}^2 p)$, and $O(n_{x} n_{y} p)$) scales quadratically with the number of samples ($n_x$, $n_y$) and linearly with the number of variables ($p$). Since the standard permutation test requires repeated re-computations of these expensive statistics it's application to large datasets can become unfeasible. While several statistical approaches have been proposed to mitigate this issue, they all sacrifice desirable statistical properties to decrease the computational cost (e.g., trade computation speed by a decrease in statistical power). A better computational strategy is to first pre-compute the Euclidean distance (kernel) matrix of the concatenated data, and then permute indexes and retrieve the corresponding elements to compute the re-sampled statistics. While this strategy can reduce the computation cost relative to the standard permutation test, it relies on the computation of a larger Euclidean distance (kernel) matrix with complexity $O((n_x + n_y)^2 p)$. In this paper, we present a novel computationally efficient permutation algorithm which only requires the pre-computation of the 3 smaller matrices and achieves large computational speedups without sacrificing finite-sample validity or statistical power. We illustrate its computational gains in a series of experiments and compare its statistical power to the current state-of-the-art approach for balancing computational cost and statistical performance. \\

\end{abstract}

\section{Introduction}

Two-sample tests based on energy distance~\cite{SzekelyRizzo2004,SzekelyRizzo2005,SzekelyRizzo2017} or maximum mean discrepancy~\cite{Gretton2006,Gretton2009,Gretton2012} statistics are commonly used statistical tests for comparing multivariate data from two groups or populations. Given two datasets, $\bfX \sim F_{\bfX}$ and $\bfY \sim F_{\bfY}$, the basic idea is to test the null hypothesis $H_0: F_{\bfX} = F_{\bfY}$, that the joint probability distribution of $\bfX$ is the same as the joint distribution of $\bfY$, against the alternative hypothesis, $H_1: F_{\bfX} \not= F_{\bfY}$, that the distributions are different. While parametric versions of two-sample tests based on maximum likelihood are available in the literature, in practice, the distributions underlying the data are almost always unknown, and non-parametric approaches based on permutation tests~\cite{Good2000} are usually preferred. In particular, permutation tests based on energy distance (ED) or maximum mean discrepancy (MMD) statistics have been widely adopted and studied in the statistics and machine learning literature~\cite{SzekelyRizzo2004,SzekelyRizzo2005,SzekelyRizzo2017,Shekhar2022}. In this paper, we focus on the ED statistic (although our contribution is directly applicable to MMD statistics too).

Permutation tests based on the ED (or MMD) statistics enjoy desirable statistical properties including finite-sample validity and consistency against all alternatives (i.e., type I error rates are controlled at the nominal significance levels, irrespective of the sample size, and the tests are able to detect differences of any arbitrary type with probability 1, as sample sizes grow to infinity)~\cite{SzekelyRizzo2004,SzekelyRizzo2005,Gretton2009,Gretton2012}. Furthermore, these tests retain their validity even in high dimensional settings where the number of variables is larger than the number of samples. (The original paper proposing the ED-test~\cite{SzekelyRizzo2004} was motivated by the high dimensional application of detecting differences between gene expression data in microarray experiments involving thousands of variables, but only a few samples.)

Despite these nice statistical properties, the ED (MMD) test statistics can be expensive to compute. For instance, in the case of the energy distance, the test statistic is given by,
\begin{equation}
\mbox{ED} = \frac{2}{n_x n_y} \sum_{k=1}^{n_x} \sum_{l=1}^{n_y} \bfD_{\bfX \bfY}^{kl} - \frac{1}{n_x^2} \sum_{k=1}^{n_x} \sum_{l=1}^{n_x} \bfD_{\bfX \bfX}^{kl} - \frac{1}{n_y^2} \sum_{k=1}^{n_y} \sum_{l=1}^{n_y} \bfD_{\bfY \bfY}^{kl}~,
\label{eq:ed.sample.estimator}
\end{equation}
where $n_x$ and $n_y$ represent, respectively, the sample sizes of the $\bfx$ and $\bfy$ datasets; $\bfD_{\bfX \bfX}$ represents the matrix of Euclidean distances among all pairs of samples of $\bfx$; $\bfD_{\bfY \bfY}$ represents the Euclidean distance matrix for all pairs of samples in $\bfy$; $\bfD_{\bfX \bfY}$ represents the matrix of Euclidean distances between all pairs of samples in $\bfx$ and $\bfy$; and $\bfD_{\bfX \bfX}^{kl}$, $\bfD_{\bfY \bfY}^{kl}$, and $\bfD_{\bfX \bfY}^{kl}$ represent the $kl$-th elements of the respective matrices. In the case of the maximum mean discrepancy, the squared test statistic (in the biased case) is given by,
\begin{equation}
\mbox{MMD}_b^2 = \frac{1}{n_x^2} \sum_{k=1}^{n_x} \sum_{l=1}^{n_x} \bfK_{\bfX \bfX}^{kl} + \frac{1}{n_y^2} \sum_{k=1}^{n_y} \sum_{l=1}^{n_y} \bfK_{\bfY \bfY}^{kl} - \frac{2}{n_x n_y} \sum_{k=1}^{n_x} \sum_{l=1}^{n_y} \bfK_{\bfX \bfY}^{kl}~,
\label{eq:mmd.biased.sample.estimator}
\end{equation}
where $\bfK_{\bfX \bfX}$, $\bfK_{\bfY \bfY}$, and $\bfK_{\bfX \bfY}$ represent the analogous kernel matrices. The issue is that the computation of each one of these 3 Euclidean distance matrices in equation \ref{eq:ed.sample.estimator} (or kernel matrices in equation \ref{eq:mmd.biased.sample.estimator}) have quadratic time complexity on the number of samples and linear time complexity on the number of variables (see Appendix A.1 for details). Since the standard permutation test requires repeated recomputations of these expensive test statistics, in practice, they can became unfeasible/impractical for tabular datasets with large number of rows (samples), and large number of columns (variables).

Several statistical approaches in the literature have been proposed to reduce computation time, including methods that approximate the test statistic null distribution, as well as, methods that leverage modified test statistics which are cheaper to compute. These computation improvements, however, come with some caveats. In one hand, methods that approximate the null distribution are either heuristic or require large samples~\cite{Gretton2009}, and do not enjoy the finite-sample validity guarantee of the permutation test. On the other hand, methods based on modified statistics such as linear-time MMD~\cite{Gretton2012}, block-MMD~\cite{Zaremba2013}, and cross-MMD~\cite{Shekhar2022} show clear computation-statistics tradeoffs~\cite{Ramdas2015}, where the decrease in computational cost comes at the expense of a decrease in statistical power.

A simple strategy for reducing the computational cost of the ED (or MMD) permutation tests is to first pre-compute the Euclidean distance (kernel) matrix associated with the concatenated dataset, $\bfW = (\bfX, \bfY)^T$, and then, at each permutation iteration, simply permute the indexes of the concatenated data and extract the respective elements of the pre-computed $\bfD_{\bfW \bfW}$ (or $\bfK_{\bfW \bfW}$) matrix in order to compute the test statistics on permuted data (as first proposed in~\cite{SzekelyRizzo2004}). While this strategy can lead to considerable reductions in computation cost, as the computation of the Euclidean distance (kernel) matrix is performed only once, the approach still requires the computation and storage of a Euclidean distance (kernel) matrix of dimension $(n_x + n_y) \times (n_x + n_y)$, which is more expensive to compute and store than the 3 Euclidean distance matrices in equation \ref{eq:ed.sample.estimator} (or kernel matrices in equation \ref{eq:mmd.biased.sample.estimator}) combined.

In this paper, we present a novel computationally efficient permutation algorithm for implementing two-sample tests based on the ED (MMD) statistics. Our proposed permutation scheme represents an improvement over the pre-computation strategy as it does not require the computation of Euclidean distance (kernel) matrices on the concatenated data matrices and, therefore, avoid the higher computation and storage requirements associated with larger datasets. Our main insights are that: (i) all the information necessary to estimate the ED (MMD) statistic in a permuted dataset is already available inside the 3 Euclidean distance matrices, $\bfD_{\bfX \bfY}$, $\bfD_{\bfX \bfX}$, and $\bfD_{\bfY \bfY}$, (or 3 kernel matrices, $\bfK_{\bfX \bfY}$, $\bfK_{\bfX \bfX}$, and $\bfK_{\bfY \bfY}$) computed on the original data (hence, there is no need for the computation of the more expensive $\bfD_{\bfW \bfW}$ or $\bfK_{\bfW \bfW}$ matrices); and (ii) that the outcome of any data permutation boils down to a swapping of elements between these 3 matrices. Hence, for any data permutation, instead of shuffling the data and recomputing the test statistic on the permuted data (which entails the re-calculation of the expensive Euclidean distance/kernel matrices at each data permutation iteration), our algorithm simply maps which elements of the 3 distance matrices are swapped for that particular data permutation. In this way, we can compute the expensive distance (kernel) matrices only once and, at each permutation iteration, compute the value of the test statistic for permuted data using the computationally much cheaper process of swapping elements among the distance (kernel) matrices and averaging the swapped matrices. Most importantly, our algorithm achieves computational gains without any sacrifice in finite-sample validity or statistical power (as it generates identical results to the standard permutation approach when evaluated on the same data permutations).

We illustrate the computational gains of our proposed algorithm relative to the standard and pre-computed permutation approaches in a series of computing time benchmarking experiments. We also compare the computation time and statistical power achieved by our efficient permutation tests against the recently proposed permutation free kernel two-sample test~\cite{Shekhar2022}, which represents the current state-of-the-art (SOTA) method in terms of the computation-statistics tradeoff, and show that our proposed method trades a small increase in computation time by a considerable increase in statistical power when compared to this SOTA baseline.

\section{Related work}

Permutation tests for the univariate two-sample problem represent a standard statistical technique (see~\cite{Good2000} and chapter 15 of~\cite{Efron1993}). Extensions to multivariate numerical variables include the ED test~\cite{SzekelyRizzo2004} and the closely related kernel-based two-sample tests based on MMD statistic~\cite{Shekhar2022}. The equivalence between (generalized) ED and MMD has been established by~\cite{Sejdinovic2013}, which showed that for any positive definite kernel adopted for the computation of the MMD, there exists a semimetric of negative type which can be used to define a generalized version of ED.

The MMD statistic has been widely used in the machine learning field. In practice, the main drawback for the use of two-sample tests based on the MMD or ED statistics is the computational coast involved in their computation. Because their complexity is quadratic on the number of samples, standard permutation tests, which require repeated computations of the statistic, can become unfeasible when the sample sizes $n_x$ and $n_y$ are large, or the data is high dimensional (large $p$). In the context of the MMD statistic, there has been several attempts in the literature to a address this issue including efforts that avoid permutations~\cite{Gretton2009,Gretton2012} and methods based on modified test statistics that are cheaper to compute~\cite{Gretton2012,Zaremba2013,Shekhar2022}.

Previous attempts that avoid permutations have been largely heuristic (i.e., lacking provable type-I error control) or have exhibited poor power. Approaches based on large deviation bounds~\cite{Gretton2006}, while distribution-free, tend to be overly conservative, exhibiting type-I errors significantly below the nominal significance level, and consequently, low power. Strategies involving approximating the MMD null distribution using Pearson curve fits or the Gamma approximation~\cite{Gretton2009} are heuristic and provide no consistency guarantees. A more principled approximation approach is based on the eigendecomposition of the Gram matrix~\cite{Gretton2009}, which is asymptotically consistent (provided some technical conditions hold). This method, however, demands considerable computational effort due to the need for a full eigendecomposition (although it can still be faster than the standard permutation approach) and provides no finite sample guarantees.

Several other methods explore the alternative strategy of modifying the test statistic to reduce computational cost and generate a more tractable limiting null distribution. One strategy involves partitioning the observations into disjoint blocks, computing the kernel-MMD statistic for each block, and then averaging the results~\cite{Zaremba2013}. While permutation tests implemented with these block-averaged approaches have finite sample validity, and can be considerably faster to compute (i.e., can scale linearly or sub-quadratically with sample size), they show a considerable reduction in power when compared to the standard permutation test. Another recent approach based on a modified test statistic is the cross-MMD test~\cite{Shekhar2022}, which leverages sample-splitting and studentization to obtain a permutation free test with quadratic time complexity. This approach, however, still trades off a large decrease in computation time (when compared to the standard permutation test) by some decrease in statistical power (although the decrease tends to be considerably less accentuated than for the block-MMD approach).

Our efficient permutation test, on the other hand, enjoys several desirable properties. First, contrary to the heuristic approaches, our test controls type I errors at the correct nominal levels. Furthermore, because our test corresponds to a permutation test, it enjoys finite-sample validity, so that the type I errors are well controlled irrespective of the sample size and, therefore, do not rely on large sample results. Second, because our test is based on the original test statistic, there is no loss of statistical power (it generates identical results to the standard permutation approach when evaluated on the same data permutations), while still achieving a considerable reduction in computation time.

\section{Background}

Throughout the text we let $\bfx$ and $\bfy$ represent the data matrices with dimensions $n_x \times p$ and $n_y \times p$ (where $n_y$ can potentially differ from $n_x$). Multidimensional datapoints (i.e., rows of the data matrices) are represented by $\bfx_i$ or $\bfy_i$.

\subsection{Distance-based two-sample permutation tests}

Let the matrix $\bfx = (\bfx_1, \ldots, \bfx_{n_x})^T$ be a sample from $F_{\bfX}$ and the matrix $\bfy = (\bfy_1, \ldots, \bfy_{n_y})^T$ be a sample from a potentially different distribution $F_{\bfY}$. Our goal is to test the null hypothesis, $H_0: F_{\bfX} = F_{\bfY}$, against the alternative hypothesis, $H_1: F_{\bfX} \not= F_{\bfY}$. Two-sample permutation tests (see Chapter 15 of~\cite{Efron1993}) provide a simple yet statistically sound way to test these hypothesis.

Let $D(\bfx, \bfy)$ be an statistic measuring a distance between the $\bfx$ and $\bfy$ datasets. For instance, $D(\bfx, \bfy)$ could be the ED, MMD, or any other statistic that measures the distance between multivariate distributions. The standard permutation test is implemented as follows: (i) concatenate $\bfx$ and $\bfy$ into a single data matrix with $n_x + n_y$ rows; (ii) sample (without replacement) $n_x$ rows from the concatenated data to create a resampled dataset $\bfx^\ast$, and use the remaining $n_y$ rows to create the resampled dataset $\bfy^\ast$; (iii) compute the statistic on the resampled datasets, $D(\bfx^\ast, \bfy^\ast)$; and (iv) repeat steps (ii) and (iii) $b$ times (where $b$ is usually a large integer). Algorithm \ref{alg:1} in Appendix A.2 describes in detail this procedure in the special case of the ED statistic.

The histogram of the $b$ resampled statistics corresponds to an estimate of the distribution of $D(\bfx, \bfy)$ under the null hypothesis $H_o: F_{\bfX} = F_{\bfY}$, and is denoted as the permutation null distribution. The permutation p-value is given by the proportion of samples in the permutation null that are greater or equal to the observed statistic, $D(\bfx, \bfy)$, computed on the original data. An unbiased estimator of the permutation p-value~\cite{Phipson2010} is given by, $(1 + \sum_{i = 1}^{b} \mathbbm{1}_{\{D(\bfx^\ast, \bfy^\ast) \geq D(\bfx, \bfy) \}})/(1 + b)$, where $\ind_{\{A\}}$ represents the indicator function assuming value 1 if event $A$ occurs and 0 otherwise.

The main drawback of this approach is that it requires the re-computation of the test statistic at each permutation iteration (see lines 15 to 17 of Algorithm \ref{alg:1}). As $b$ is usually set to values between 100 and 1000, we have that the computationally heavy calculations of the distance matrices are repeated a large number of times.

\subsection{Pre-computation approach}

A simple strategy for reducing the computational cost of the standard permutation test (originally proposed in~\cite{SzekelyRizzo2004}) is to first pre-compute the Euclidean distance (kernel) matrices using the concatenated data, $\bfw = (\bfx, \bfy)^T$, and then, at each permutation iteration, compute the test statistic on permuted data by simply permuting the indexes of the concatenated data and extracting the respective elements from the pre-computed matrix, rather than recomputing the matrices on permuted data. Algorithm \ref{alg:pre} in Appendix A.2 describes this strategy for the energy distance statistic.

Because the Euclidean distance (kernel) matrix of the concatenated data, $\bfD_{\bfW \bfW}$ ($\bfK_{\bfW \bfW}$), only needs to be computed once, this strategy can lead to considerable speedups in the computation of the permutation test. However, it is important to point out that the time complexity of the $\bfD_{\bfW \bfW}$ ($\bfK_{\bfW \bfW}$) matrix computation is of order $O((n_x + n_y)^2 p)$ and, therefore, higher than the combined complexity of the 3 separate Euclidean distance (kernel) matrices calculations necessary for the estimation of the ED (MMD) statistics (whose complexities are given, respectively, by $O(n_x^2 p)$, $O(n_y^2 p)$, and $O(n_x n_y p)$). Additionally, the space complexity of this pre-computation strategy, $O((n_x + n_y)^2)$, is also higher than the combined complexity of the 3 separate matrix calculations (given, respectively, by $O(n_x^2)$, $O(n_y^2)$, and $O(n_x n_y)$).

\section{Our contribution}

\subsection{The proposed computationally efficient permutation test}

In the two-sample problem, a permutation of the data is obtained by resampling rows from the concatenated data matrix. As a toy example, suppose that,
\begin{equation}
\bfx = \left(\hspace{-0.1cm}
\begin{array}{c}
\bfx_1 \\
\bfx_2 \\
\bfx_3 \\
\bfx_4 \\
\bfx_5 \\
\end{array}
\hspace{-0.1cm} \right),
\hspace{0.5cm}
\bfy = \left(\hspace{-0.1cm}
\begin{array}{c}
\bfy_1 \\
\bfy_2 \\
\bfy_3 \\
\bfy_4 \\
\end{array}
\hspace{-0.1cm} \right).
\label{eq:original.data}
\end{equation}
Then, one possible permutation of the data is given, for example, by,
\begin{equation}
\bfax = \left(\hspace{-0.1cm}
\begin{array}{c}
\bfy_2 \\
\bfx_4 \\
\bfx_5 \\
\bfy_1 \\
\bfx_2 \\
\end{array}
\hspace{-0.1cm} \right),
\hspace{0.3cm}
\bfay = \left(\hspace{-0.1cm}
\begin{array}{c}
\bfx_1 \\
\bfx_3 \\
\bfy_3 \\
\bfy_4 \\
\end{array}
\hspace{-0.1cm} \right).
\label{eq:permuted.data}
\end{equation}

As briefly mentioned in the Introduction, the main insights that allow a more efficient implementation of two-sample permutation tests for the ED (MMD) statistic are that:
\begin{enumerate}
\item All the information necessary to estimate the ED (MMD) statistic in a permuted dataset is already available inside the 3 Euclidean distance (kernel) matrices computed on the original data.
\item The outcome of any data permutation boils down to a swapping of elements between these 3 matrices.
\end{enumerate}

To illustrate these points, let's consider the ED statistic. The Euclidean distance matrices computed in the original toy data in (\ref{eq:original.data}) are given by,
\begin{align}
&\bfD_{\bfX \bfX} = \left(\hspace{-0.2cm}
\begin{array}{ccccc}
D(\bfx_1,\bfx_1) & D(\bfx_1,\bfx_2) & D(\bfx_1,\bfx_3) & D(\bfx_1,\bfx_4) & D(\bfx_1,\bfx_5) \\
D(\bfx_2,\bfx_1) & D(\bfx_2,\bfx_2) & D(\bfx_2,\bfx_3) & D(\bfx_2,\bfx_4) & D(\bfx_2,\bfx_5) \\
D(\bfx_3,\bfx_1) & D(\bfx_3,\bfx_2) & D(\bfx_3,\bfx_3) & D(\bfx_3,\bfx_4) & D(\bfx_3,\bfx_5) \\
D(\bfx_4,\bfx_1) & D(\bfx_4,\bfx_2) & D(\bfx_4,\bfx_3) & D(\bfx_4,\bfx_4) & D(\bfx_4,\bfx_5) \\
D(\bfx_5,\bfx_1) & D(\bfx_5,\bfx_2) & D(\bfx_5,\bfx_3) & D(\bfx_5,\bfx_4) & D(\bfx_5,\bfx_5) \\
\end{array}
\hspace{-0.2cm} \right),
\label{eq:euclid.matrix.x.x}
\end{align}
\begin{equation}
\bfD_{\bfY \bfY} = \left(\hspace{-0.2cm}
\begin{array}{cccc}
D(\bfy_1,\bfy_1) & D(\bfy_1,\bfy_2) & D(\bfy_1,\bfy_3) & D(\bfy_1,\bfy_4) \\
D(\bfy_2,\bfy_1) & D(\bfy_2,\bfy_2) & D(\bfy_2,\bfy_3) & D(\bfy_2,\bfy_4) \\
D(\bfy_3,\bfy_1) & D(\bfy_3,\bfy_2) & D(\bfy_3,\bfy_3) & D(\bfy_3,\bfy_4) \\
D(\bfy_4,\bfy_1) & D(\bfy_4,\bfy_2) & D(\bfy_4,\bfy_3) & D(\bfy_4,\bfy_4) \\
\end{array}
\hspace{-0.2cm} \right),
\label{eq:euclid.matrix.y.y}
\end{equation}
\begin{equation}
\bfD_{\bfX \bfY} = \left(\hspace{-0.2cm}
\begin{array}{ccccc}
D(\bfx_1,\bfy_1) & D(\bfx_1,\bfy_2) & D(\bfx_1,\bfy_3) & D(\bfx_1,\bfy_4) \\
D(\bfx_2,\bfy_1) & D(\bfx_2,\bfy_2) & D(\bfx_2,\bfy_3) & D(\bfx_2,\bfy_4) \\
D(\bfx_3,\bfy_1) & D(\bfx_3,\bfy_2) & D(\bfx_3,\bfy_3) & D(\bfx_3,\bfy_4) \\
D(\bfx_4,\bfy_1) & D(\bfx_4,\bfy_2) & D(\bfx_4,\bfy_3) & D(\bfx_4,\bfy_4) \\
D(\bfx_5,\bfy_1) & D(\bfx_5,\bfy_2) & D(\bfx_5,\bfy_3) & D(\bfx_5,\bfy_4) \\
\end{array}
\hspace{-0.2cm} \right),
\label{eq:euclid.matrix.x.y}
\end{equation}
whereas the Euclidean distance matrices computed on the permuted data in (\ref{eq:permuted.data}) are,
\begin{align}
&\bfD_{\bfaX \bfaX} = \left(\hspace{-0.2cm}
\begin{array}{ccccc}
D(\bfy_2,\bfy_2) & D(\bfy_2,\bfx_4) & D(\bfy_2,\bfx_5) & D(\bfy_2,\bfy_1) & D(\bfy_2,\bfx_2) \\
D(\bfx_4,\bfy_2) & D(\bfx_4,\bfx_4) & D(\bfx_4,\bfx_5) & D(\bfx_4,\bfy_1) & D(\bfx_4,\bfx_2) \\
D(\bfx_5,\bfy_2) & D(\bfx_5,\bfx_4) & D(\bfx_5,\bfx_5) & D(\bfx_5,\bfy_1) & D(\bfx_5,\bfx_2) \\
D(\bfy_1,\bfy_2) & D(\bfy_1,\bfx_4) & D(\bfy_1,\bfx_5) & D(\bfy_1,\bfy_1) & D(\bfy_1,\bfx_2) \\
D(\bfx_2,\bfy_2) & D(\bfx_2,\bfx_4) & D(\bfx_2,\bfx_5) & D(\bfx_2,\bfy_1) & D(\bfx_2,\bfx_2) \\
\end{array}
\hspace{-0.2cm} \right),
\label{eq:euclid.matrix.xstar.xstar}
\end{align}
\begin{equation}
\bfD_{\bfaY \bfaY} = \left(\hspace{-0.2cm}
\begin{array}{cccc}
D(\bfx_1,\bfx_1) & D(\bfx_1,\bfx_3) & D(\bfx_1,\bfy_3) & D(\bfx_1,\bfy_4) \\
D(\bfx_3,\bfx_1) & D(\bfx_3,\bfx_3) & D(\bfx_3,\bfy_3) & D(\bfx_3,\bfy_4) \\
D(\bfy_3,\bfx_1) & D(\bfy_3,\bfx_3) & D(\bfy_3,\bfy_3) & D(\bfy_3,\bfy_4) \\
D(\bfy_4,\bfx_1) & D(\bfy_4,\bfx_3) & D(\bfy_4,\bfy_3) & D(\bfy_4,\bfy_4) \\
\end{array}
\hspace{-0.2cm} \right),
\label{eq:euclid.matrix.ystar.ystar}
\end{equation}
\begin{equation}
\bfD_{\bfaX \bfaY} = \left(\hspace{-0.2cm}
\begin{array}{ccccc}
D(\bfy_2,\bfx_1) & D(\bfy_2,\bfx_3) & D(\bfy_2,\bfy_3) & D(\bfy_2,\bfy_4) \\
D(\bfx_4,\bfx_1) & D(\bfx_4,\bfx_3) & D(\bfx_4,\bfy_3) & D(\bfx_4,\bfy_4) \\
D(\bfx_5,\bfx_1) & D(\bfx_5,\bfx_3) & D(\bfx_5,\bfy_3) & D(\bfx_5,\bfy_4) \\
D(\bfy_1,\bfx_1) & D(\bfy_1,\bfx_3) & D(\bfy_1,\bfy_3) & D(\bfy_1,\bfy_4) \\
D(\bfx_2,\bfx_1) & D(\bfx_2,\bfx_3) & D(\bfx_2,\bfy_3) & D(\bfx_2,\bfy_4) \\
\end{array}
\hspace{-0.2cm} \right),
\label{eq:euclid.matrix.xstar.ystar}
\end{equation}
Direct comparison of these matrices shows that all the elements in the permuted distance matrices $\bfD_{\bfaX \bfaX}$, $\bfD_{\bfaY \bfaY}$, and $\bfD_{\bfaX \bfaY}$ were already computed in the original data distance matrices (illustrating point (1), that all the pairwise distances necessary for computing the ED statistic in the permutated data are already available in the original Euclidean distance matrices, $\bfD_{\bfX \bfX}$, $\bfD_{\bfY \bfY}$, and $\bfD_{\bfX \bfY}$). Furthermore, this example clearly illustrates point (2), that the result of a data permutation reduces to a swapping of elements between the 3 original distance matrices. (For instance, because the elements of $\bfax$ now contain samples originally from both $\bfx$ and $\bfy$, we have that $\bfD_{\bfaX \bfaX}$ contain elements that belonged originally to $\bfD_{\bfX \bfX}$, $\bfD_{\bfY \bfY}$, and $\bfD_{\bfX \bfY}$.)

In Algorithms \ref{alg:eff} and \ref{alg:eff2} we describe how to leverage these insights to calculate the values of the ED statistic in permuted data, without having to recompute the Euclidean distances for each new data permutation (or having to work with the larger pre-computed $\bfD_{\bfW \bfW}$ matrix, derived from the concatenated data). In particular, Algorithm \ref{alg:eff2} (which is called from inside Algorithm \ref{alg:eff}) describes how to obtain the mapping between a data permutation and the positions of the elements of the Euclidean distance matrices computed on the original data.

\begin{algorithm}[!h]
\caption{Efficient permutation two-sample test for the energy distance statistic}\label{alg:eff}
\KwData{Datasets $\bfx$ and $\bfy$; and number of permutations, $b$}
\ShowLn $\bfD_{\bfX \bfX} \leftarrow \mbox{EuclideanDistanceMatrix}(\bfx, \bfx)$ \\
\ShowLn $\bfD_{\bfY \bfY} \leftarrow \mbox{EuclideanDistanceMatrix}(\bfy, \bfy)$ \\
\ShowLn $\bfD_{\bfX \bfY} \leftarrow \mbox{EuclideanDistanceMatrix}(\bfx, \bfy)$ \\
\ShowLn $\bfD_{\bfY \bfX} \leftarrow \bfD_{\bfX \bfY}^T$ \\
\ShowLn $n_x \leftarrow \mbox{NumberOfRows}(\bfx)$ \\
\ShowLn $n_y \leftarrow \mbox{NumberOfRows}(\bfy)$ \\
\ShowLn $\mbox{ED}_{o} \leftarrow \frac{2}{n_x n_y} \sum_{k=1}^{n_x} \sum_{l=1}^{n_y} \bfD_{\bfX \bfY}^{kl} - \frac{1}{n_x^2} \sum_{k=1}^{n_x} \sum_{l=1}^{n_x} \bfD_{\bfX \bfX}^{kl} - \frac{1}{n_y^2} \sum_{k=1}^{n_y} \sum_{l=1}^{n_y} \bfD_{\bfY \bfY}^{kl}$ \\
\ShowLn $\mbox{ED}^\ast \leftarrow [\;]$ \\
\ShowLn \For{$i$ in 1, \ldots, b} {
  \ShowLn $[\, \bfi_1 \, , \, \bfi_2 \, , \, \bfai_1 \, , \, \bfai_2 \, , \, \bfj_1 \, , \, \bfj_2 \, , \, \bfaj_1 \, , \, \bfaj_2 \,] \leftarrow \mbox{PermutationIndexesTwoSampleTest}(n_x, n_y)$ \\
  \ShowLn $\bfD_{\bfaX \bfaX} \leftarrow [ \, , \,]$ \\
  \ShowLn $\bfD_{\bfaX \bfaX}[\, \bfai_1 \, , \, \bfai_1 \,] \leftarrow \bfD_{\bfX \bfX}[\, \bfi_1 \, , \, \bfi_1 \,]$ \\
  \ShowLn $\bfD_{\bfaX \bfaX}[\, \bfai_1 \, , \, \bfai_2 \,] \leftarrow \bfD_{\bfX \bfY}[\, \bfi_1 \, , \, \bfi_2 \,]$ \\
  \ShowLn $\bfD_{\bfaX \bfaX}[\, \bfai_2 \, , \, \bfai_1 \,] \leftarrow \bfD_{\bfY \bfX}[\, \bfi_2 \, , \, \bfi_1 \,]$ \\
  \ShowLn $\bfD_{\bfaX \bfaX}[\, \bfai_2 \, , \, \bfai_2 \,] \leftarrow \bfD_{\bfY \bfY}[\, \bfi_2 \, , \, \bfi_2 \,]$ \\
  \ShowLn $\bfD_{\bfaY \bfaY} \leftarrow [ \, , \,]$ \\
  \ShowLn $\bfD_{\bfaY \bfaY}[\, \bfaj_1 \, , \, \bfaj_1 \,] \leftarrow \bfD_{\bfX \bfX}[\, \bfj_1 \, , \, \bfj_1 \,]$ \\
  \ShowLn $\bfD_{\bfaY \bfaY}[\, \bfaj_1 \, , \, \bfaj_2 \,] \leftarrow \bfD_{\bfX \bfY}[\, \bfj_1 \, , \, \bfj_2 \,]$ \\
  \ShowLn $\bfD_{\bfaY \bfaY}[\, \bfaj_2 \, , \, \bfaj_1 \,] \leftarrow \bfD_{\bfY \bfX}[\, \bfj_2 \, , \, \bfj_1 \,]$ \\
  \ShowLn $\bfD_{\bfaY \bfaY}[\, \bfaj_2 \, , \, \bfaj_2 \,] \leftarrow \bfD_{\bfY \bfY}[\, \bfj_2 \, , \, \bfj_2 \,]$ \\
  \ShowLn $\bfD_{\bfaX \bfaY} \leftarrow [ \, , \,]$ \\
  \ShowLn $\bfD_{\bfaX \bfaY}[\, \bfai_1 \, , \, \bfaj_1 \,] \leftarrow \bfD_{\bfX \bfX}[\, \bfi_1 \, , \, \bfj_1 \,]$ \\
  \ShowLn $\bfD_{\bfaX \bfaY}[\, \bfai_1 \, , \, \bfaj_2 \,] \leftarrow \bfD_{\bfX \bfY}[\, \bfi_1 \, , \, \bfj_2 \,]$ \\
  \ShowLn $\bfD_{\bfaX \bfaY}[\, \bfai_2 \, , \, \bfaj_1 \,] \leftarrow \bfD_{\bfY \bfX}[\, \bfi_2 \, , \, \bfj_1 \,]$ \\
  \ShowLn $\bfD_{\bfaX \bfaY}[\, \bfai_2 \, , \, \bfaj_2 \,] \leftarrow \bfD_{\bfY \bfY}[\, \bfi_2 \, , \, \bfj_2 \,]$ \\
  \ShowLn $\mbox{ED}^\ast[i] \leftarrow \frac{2}{n_x n_y} \sum_{k=1}^{n_x} \sum_{l=1}^{n_y} \bfD_{\bfaX \bfaY}^{kl} - \frac{1}{n_x^2} \sum_{k=1}^{n_x} \sum_{l=1}^{n_x} \bfD_{\bfaX \bfaX}^{kl} - \frac{1}{n_y^2} \sum_{k=1}^{n_y} \sum_{l=1}^{n_y} \bfD_{\bfaY \bfaY}^{kl}$ \\
}
\ShowLn $\mbox{perm. p-value} \leftarrow (1 + \sum_{i=1}^{b} \ind \{ \mbox{ED}^\ast[i] \ge \mbox{ED}_o \})/(b + 1)$ \\
\KwResult{Return the permutation p-value, and permutation null distribution (ED$^\ast$)}
\end{algorithm}

\begin{algorithm}[!h]
\caption{PermutationIndexesTwoSampleTest$(n_x, n_y)$}\label{alg:eff2}
\KwData{Number of samples on group 1 ($n_x$), and on group 2 ($n_y$)}
\ShowLn $n \leftarrow n_x + n_y$ \\
\ShowLn $idx_w \leftarrow (1, \ldots, n)$ \\
\ShowLn $idx_g \leftarrow (1, \ldots, n_x, 1, \ldots, n_y)$ \\
\ShowLn $idx_{p1} \leftarrow \mbox{SampleWithoutReplacement}(idx_w, n_x)$ \\
\ShowLn $idx_{p2} \leftarrow \mbox{SetDifference}(idx_w, idx_{p1})$ \\
\ShowLn $\bfi_1 \leftarrow idx_g[idx_{p1}[\mbox{Which}(idx_{p1} \le n_x)]]$ \\
\ShowLn $\bfai_1 \leftarrow (1, \ldots, \mbox{Length}(i_1))$ \\
\ShowLn $\bfi_2 \leftarrow idx_g[idx_{p1}[\mbox{Which}(idx_{p1} > n_x)]]$ \\
\ShowLn $\bfai_2 \leftarrow (1 + \mbox{Length}(\bfi_1), \ldots, n_x)$ \\
\ShowLn $\bfj_1 \leftarrow idx_g[idx_{p2}[\mbox{Which}(idx_{p2} \le n_x)]]$ \\
\ShowLn $\bfaj_1 \leftarrow (1, \ldots, \mbox{Length}(\bfj_1))$ \\
\ShowLn $\bfj_2 \leftarrow idx_g[idx_{p2}[\mbox{Which}(idx_{p2} > n_x)]]$ \\
\ShowLn $\bfaj_2 \leftarrow (1 + \mbox{Length}(\bfj_1), \ldots, n_y)$ \\
\KwResult{Return the list of indexes: $[\, \bfi_1 \, , \, \bfi_2 \, , \, \bfai_1 \, , \, \bfai_2 \, , \, \bfj_1 \, , \, \bfj_2 \, , \, \bfaj_1 \, , \, \bfaj_2 \,]$}
\end{algorithm}

\subsubsection{Description of Algorithm 1}

Starting with Algorithm \ref{alg:eff}, note that lines 1 to 3 compute the Euclidean distance matrices on the original data (and that those are only computed once), while lines 5 to 7 describe the computation of the ED statistic on the original data.

Lines 8 to 26 describe the computation of the permutation null distribution. Line 8 creates an empty vector for storing the ED statistics computed on each of the $b$ data permutations. At each iteration $i$ of the for-loop, the first step (described in line 10, which calls Algorithm \ref{alg:eff2}) is to perform a permutation and return:
\begin{itemize}
\item The indexes of the elements of the original distance matrices, $\bfD_{\bfX \bfX}$, $\bfD_{\bfY \bfY}$, and $\bfD_{\bfX \bfY}$, that need to be swapped in order to generate the corresponding distance distance matrices in the permuted dataset ($\bfD_{\bfaX \bfaX}$, $\bfD_{\bfaY \bfaY}$, and $\bfD_{\bfaX \bfaY}$), where:
    \begin{itemize}
    \item $\bfi_1$: corresponds to the indexes of the samples of $\bfx$ that where assigned to $\bfax$ during the permutation process.
    \item $\bfi_2$: corresponds to the indexes of the samples of $\bfy$ that where assigned to $\bfax$ during the permutation process.
    \item $\bfj_1$: corresponds to the indexes of the samples of $\bfx$ that where assigned to $\bfay$ during the permutation process.
    \item $\bfj_2$: corresponds to the indexes of the samples of $\bfy$ that where assigned to $\bfay$ during the permutation process.
    \end{itemize}
\item The corresponding indexes ($\bfai_1$, $\bfai_2$, $\bfaj_1$, and $\bfaj_2$) of the permuted distance matrices (i.e., the positions of the permuted distance matrices which will store the swapped elements from the original distance matrices).
\end{itemize}

Once these indexes are determined, all that is left to do is to fill in permuted distance matrices with the swapped elements from the original distance matrices. Lines 11 to 15 describe this process for the $\bfD_{\bfaX \bfaX}$ matrix, while lines 16 to 20 and lines 21 to 25 describe this process for the $\bfD_{\bfaY \bfaY}$ and $\bfD_{\bfaX \bfaY}$ matrices, respectively. Finally, line 26 describes the computation of the ED statistic for the permuted data, and line 27 describes the computation of the permutation p-value.

Observe as well that a memory more efficient version of Algorithm \ref{alg:eff} (which avoids having to store the $\bfD_{\bfY \bfX}$ matrix) can be obtained by removing line 4 and replacing: $\bfD_{\bfY \bfX}[\bfi_2 , \bfi_1]$ by $\bfD_{\bfX \bfY}[\bfi_1 , \bfi_2]^T$ in line 14; $\bfD_{\bfY \bfX}[\bfj_2 , \bfj_1]$ by $\bfD_{\bfX \bfY}[\bfj_1 , \bfj_2]^T$ in line 19; and $\bfD_{\bfY \bfX}[\bfi_2 , \bfj_1]$ by $\bfD_{\bfX \bfY}[\bfj_1 , \bfi_2]^T$ in line 24.

\subsubsection{Description of Algorithm 2}

Now we describe Algorithm \ref{alg:eff2}. Line 2 creates a vector of indexes ranging from 1 to $n = n_x + n_y$, denoted $idx_w$, which corresponds to the indexes of the dataset $\bfw$ obtained by concatenating the $\bfx$ and $\bfy$ datasets.

Line 3 creates an alternative vector of length $n$, denoted $idx_g$, by concatenating the vectors of indexes $(1, \ldots, n_x)$ and the vector of indexes $(1, \ldots, n_y)$. (This vector will be used to map the indexes of the concatenated data to the indexes of the original datasets $\bfx$ and $\bfy$ and, by extension, to the indexes of the Euclidean distance matrices.)

Lines 4 and 5 perform a permutation of the indexes of the concatenated dataset $\bfw$ in order to obtain the indexes of the samples assigned to the permuted datasets $\bfax$ and $\bfay$. Note that in line 4 the algorithm randomly samples $n_x$ elements from the $idx_w$ vector (which correspond to the indexes of the samples of the concatenated data that are assigned to $\bfax$ during the permutation process). In Line 5, the algorithm obtains the set of remaining indexes (which correspond to the indexes of $\bfw$ assigned to $\bfay$ during the permutation process) by taking the set difference between $idx_w$ and $idx_{p1}$.

Line 6 finds the indexes of the samples from the $\bfx$ dataset that were assigned to $\bfax$ during the permutation process. Note that because the concatenated data $\bfw$ is obtained by appending $\bfy$ to the end of $\bfx$, we have that any element of $idx_{p1}$ which is less or equal to $n_x$ corresponds to a sample from $\bfx$. Hence, the operation $\mbox{Which}(idx_{p1} \le n_x)$ is used to find the positions of $idx_{p1}$ which contain these elements, and the operation $idx_{p1}[\mbox{Which}(idx_{p1} \le n_x)]$ is used to find the indexes of the samples of $\bfx$ that were assigned to $\bfax$ relative to the concatenated data $\bfw$ (indexed by $idx_w$). Finally, the operation $idx_g[idx_{p1}[\mbox{Which}(idx_{p1} \le n_x)]]$ is used to map back these indexes to the indexes of the original $\bfx$ data (and, consequently, to the row indexes of $\bfD_{\bfX, \bfY}$, column indexes of $\bfD_{\bfY, \bfX}$, and row and column indexes of $\bfD_{\bfX, \bfX}$). Line 7 generates the corresponding indexes, $\bfai_1$, of the permuted data Euclidean distance matrices that will receive the elements of the original Euclidean distance matrices indexed by $\bfi_1$ (note that $\bfai_1$ has the same length as $\bfi_1$.)

Similarly, line 8 finds the indexes of the samples from the $\bfy$ dataset that were assigned to $\bfax$ during the permutation process. Since any element of $idx_{p1}$ which is greater to $n_x$ corresponds to a sample from $\bfy$, the operation $idx_{p1}[\mbox{Which}(idx_{p1} > n_x)]$ is used to find the indexes of the samples from $\bfy$ which were assigned to $\bfax$ relative to $\bfw$, and the operation $idx_g[idx_{p1}[\mbox{Which}(idx_{p1} > n_x)]]$ is used to map back these indexes to the indexes of the original $\bfy$ data (and, consequently, to the row indexes of $\bfD_{\bfY, \bfX}$, column indexes of $\bfD_{\bfX, \bfY}$, and row and column indexes of $\bfD_{\bfY, \bfY}$). Line 9 generates corresponding indexes, $\bfai_2$, for the the permuted data Euclidean distance matrices (note that, again, $\bfai_2$ has the same length as $\bfi_2$.)

Finally, lines 10 and 12 describe the analogous operations to obtain the indexes from samples of the $\bfx$ and $\bfy$ datasets which were assigned to $\bfay$ during the permutation process (and, consequently, the indexes of the elements of the original Euclidean distance matrices that are swapped), while lines 11 and 13 generate the corresponding indexes for the permuted distance matrices.

For the sake of clarity, we also provide in Appendix A.4 a step-by-step illustration of the application of Algorithms \ref{alg:eff} and \ref{alg:eff2} to the permutation example in equation (\ref{eq:permuted.data}).

\section{Experiments and illustrations}

In this section, we present experiments evaluating the statistical and computational performance of the efficient permutation ED-test.

\subsection{Computation time comparisons}

We performed two computing time benchmarking experiments comparing the standard test (Algorithm \ref{alg:1}), pre-computed test (Algorithm \ref{alg:pre}), and efficient permutation ED test (Algorithms \ref{alg:eff} and \ref{alg:eff2}) against the cross-ED test (described in Algorithm \ref{alg:xed} in Appendix A.3). In the first experiment, we investigate compute time for datasets containing a fixed number of variables $p = 500$, but with number of samples, $n_x = n_y$, varying from 100 to 1000 (at increments of 100). In the second, we investigate compute time for datasets with a fixed number of samples $n_x = n_y = 200$, but with number of variables varying from 200 to 2000 (at increments of 200). In all experiments, we simulate $\bfx$ and $\bfy$ datasets from $N_p(\bfzero, \bfI_p)$ distributions, and report the user time measured by the \texttt{system.time} function of the R~\cite{rproject2019} base distribution. All experiments were performed on a Windows machine with processor Intel(R) Core(TM) i7-7820HQ CPU \@ 2.90GHz 2.90 GHz and 64 GB of RAM. Each experiment was replicated 100 times (with the exception of the standard permutation test that was replicated only 10 times). In all experiments, the number of permutations was set to 200.

\begin{figure}[!h]
\centerline{\includegraphics[width=3.8in]{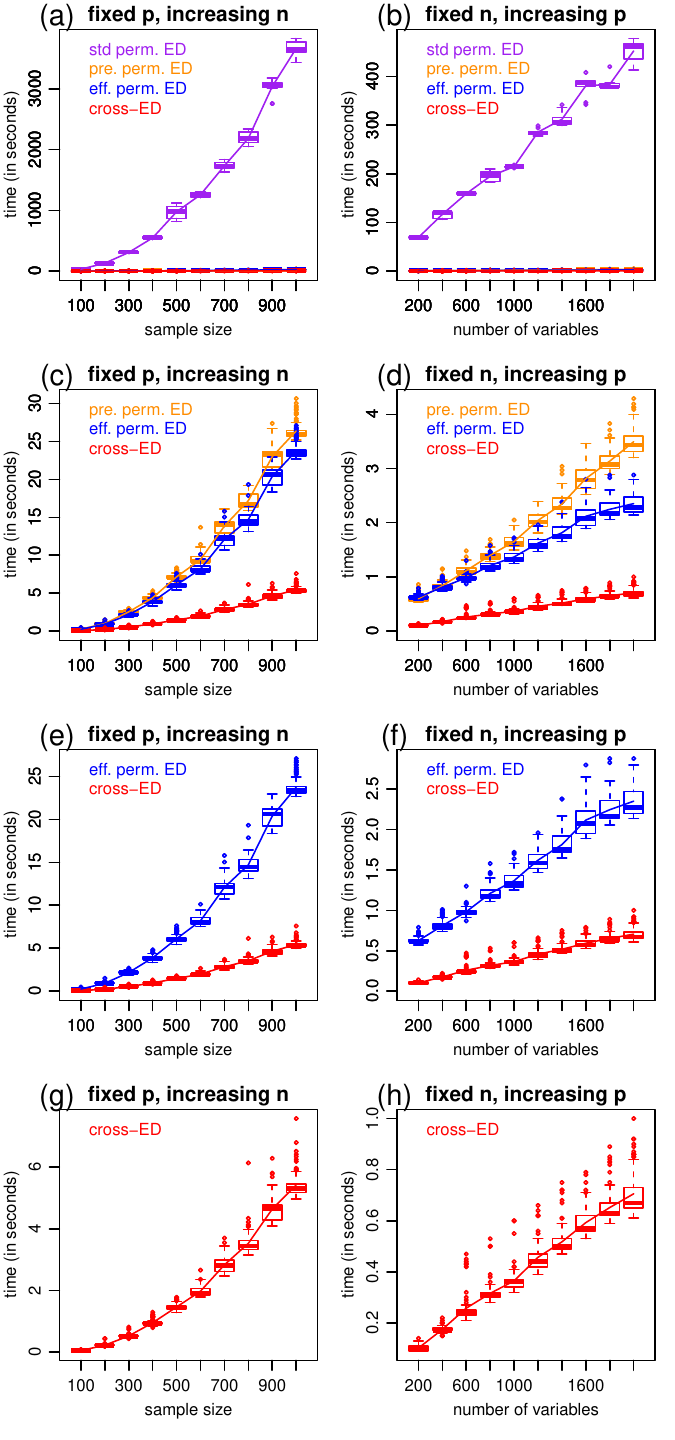}}
\vskip -0.2in
\caption{Compute time benchmarking experiments.}
\label{fig:time.comparisons}
\end{figure}

Figures \ref{fig:time.comparisons}a and b report the computation time (in seconds) for the 4 tests based on the first and second experiments, respectively. These results illustrate the high speed ups obtained by the cross-ED (red), pre-computed perm. ED (orange) and efficient perm. ED (blue) tests in comparison with the standard perm. ED test (purple). In both panels, the boxplots report the results from the experiment replications, while the solid lines represent the averages across the replications.

Figures \ref{fig:time.comparisons}c and d present the same comparisons after the removal of the standard perm. test results. These comparisons show that the efficient perm. approach tends to be slightly faster than the pre-computation approach, illustrating that the computation of a single Euclidean matrix of dimension $(n_x + n_y) \times (n_x + n_y)$ was still more expensive than the computation of 3 smaller matrices of dimension $n_x \times n_x$, $n_y \times n_y$, and $n_x \times n_y$ (plus the additional computations performed by Algorithm \ref{alg:eff2}). Figure \ref{fig:time.comparisons}d also shows that the computing time differences tend to be greater for larger values of $p$.

Figures \ref{fig:time.comparisons}e and f compare only the efficient permutation and cross-ED tests, and illustrate that while the cross-ED test (red) is faster than the efficient permutation test (blue), the small dynamic ranges of the y-axes of both panels (when compared to the respective ranges in panels a and b) show that the absolute difference in computation time is still rather small. For both methods, the bulk of the speed-up is realized by avoiding repeated re-computations of the ED statistic, as illustrated in panels a and b.

Finally, Figures \ref{fig:time.comparisons}g and h report the computation time for the cross-ED test alone. The greater computational efficiency of the cross-ED test results from the fact that it only relies on the computation of 4 distinct Euclidean distance matrices based on split datasets with roughly half of the number of samples of the original datasets, which are computed only once, as described in Algorithm \ref{alg:xed}. Explicitly, the time complexity of each Euclidean distance matrix computation on lines 9 to 12 of Algorithm \ref{alg:xed} is given, respectively, by $O(n_{x_1} n_{x_2} p)$, $O(n_{y_1} n_{y_2} p)$, $O(n_{x_1} n_{y_2} p)$, and $O(n_{y_1} n_{x_2} p)$, where $n_{x_1} = \lfloor n_x/2 \rfloor$, $n_{x_2} = \lceil n_x/2 \rceil$, $n_{y_1} = \lfloor n_y/2 \rfloor$, and $n_{y_2} = \lceil n_y/2 \rceil$ (and $\lfloor z \rfloor$ represents the greatest integer less than or equal to $z$, while $\lceil z \rceil$ represents the least integer greater or equal to $z$). The computational complexity of the efficient permutation ED-test, on the other hand, is dominated by the computation of the 3 Euclidean distance matrices in lines 1, 2, and, 3 of Algorithm \ref{alg:eff}, with time complexity of order $O(n_{x}^2 p)$, $O(n_{y}^2 p)$, and $O(n_{x} n_{y} p)$, respectively.

As expected, for all tests, we observed an approximately quadratic increase in computation time in the first experiment, where $p$ is kept fixed while $n_x = n_y$ is allowed to increase (Figure \ref{fig:time.comparisons}a, c, e and g), but only an approximately linear increase in the second experiment, where $n_x = n_y$ is kept fixed while $p$ is allowed to increase (Figure \ref{fig:time.comparisons}b, d, f and h).

\subsection{Statistical performance evaluations}

\begin{figure}[!h]
\centerline{\includegraphics[width=4in]{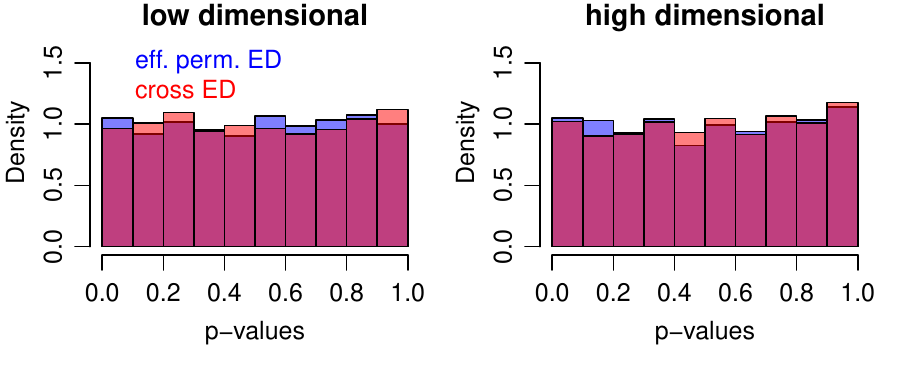}}
\vskip -0.2in
\caption{P-value distributions for data generated under the null hypothesis.}
\label{fig:null.distr.synth}
\end{figure}

Here we present statistical performance comparisons between the efficient permutation ED test and the cross-ED test. First, to illustrate that both tests have finite sample validity we performed 2 experiments where we generate data under the null hypothesis that $F_{\bfX} = F_{\bfY}$, by sampling both $\bfX$ and $\bfY$ datasets from a $N_p(\bfzero, \bfI_p)$ distribution. The first experiment, involves a low dimensional setting where $n = 200$ and $p = 100$. The second, involves a high dimensional setting where $n = 200$ and $p = 1000$. Each experiment was replicated 2000 times and Figure \ref{fig:null.distr.synth} reports the distributions of the p-values generated by the efficient permutation ED-test (blue) and the cross-ED test (red). For both tests, in both experiments, the distribution of the p-values is uniform illustrating that type I error rates are controlled at the nominal significance levels.

Second, to compare the statistical power of these two tests, we closely follow the experimental setup adopted in~\cite{Shekhar2022} and generate data under the alternative hypothesis that $F_{\bfX} \not= F_{\bfY}$ by sampling $\bfX$ and $\bfY$ from $\bfX \sim  N_p(\bfzero, \bfI_p)$ and $\bfY \sim N_p(\bfmu_{\epsilon,j} , \bfI_p)$, where the mean vector $\bfmu_{\epsilon,j}$ is obtained by setting the first $j \le p$ coordinates of $\bfzero$ to $\epsilon$. We perform 3 sets of experiments. Setting 1 corresponds to a low dimensional setting where we perform 16 experiments over increasing values of $n_x = n_y$ (ranging from 60 to 360 at increments of 20) while keeping $p$ fixed at 50, and setting $j = 5$ and $\epsilon = 0.25$. Settings 2 and 3 correspond to a high dimensional setting where we perform 19 experiments over increasing values of $p$ (ranging from 200 to 2000 at increments of 100), while keeping $n_x = n_y$ fixed at 200, and $\epsilon = 0.1$. The difference between settings 2 and 3 is that in setting 2 we fix $j = 50$, while in setting 3 we let $j = 0.1 p$ (so that the proportion of non-zero entries was fixed at 10\% across all 19 experiments). Each experiment was replicated 500 times and the empirical power (at a significance level $\alpha = 0.05$) was computed as the proportion of replications which produced p-values lower than 0.05. In all experiments, the permutation test was based on 200 permutations.

\begin{figure}[!h]
\centerline{\includegraphics[width=3.8in]{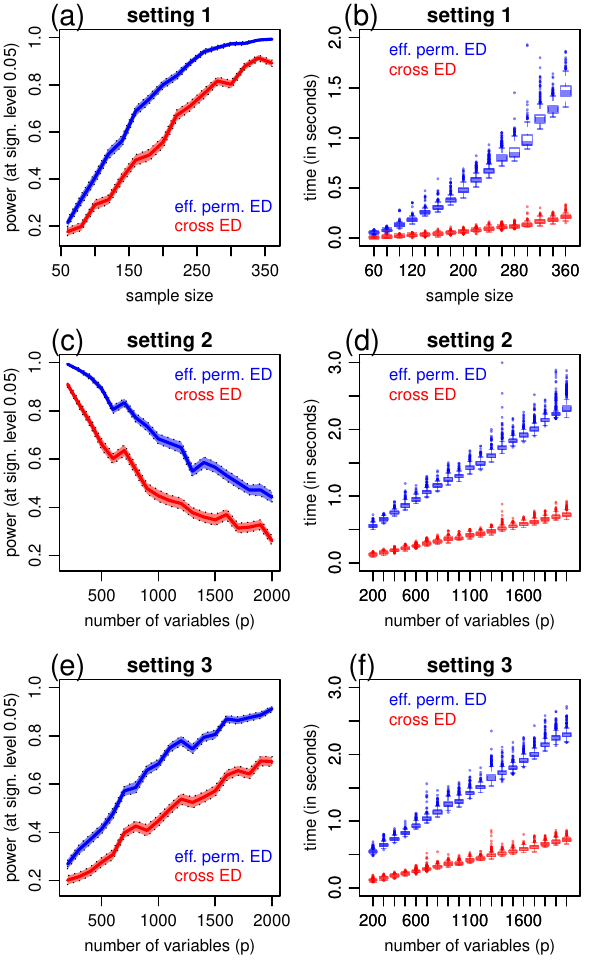}}
\vskip -0.2in
\caption{Experimental results. Setting 1 corresponds to a low-dimensional setting where $p < n_x$, while settings 2 and 3 correspond to high-dimensional settings with $p > n_x$. The difference between settings 2 and 3 is that in setting 2 we fix the number of non-zero entries of $\bfmu_{\epsilon,j}$ at $j = 50$, while in setting 3 the proportion of non-zero entries is kept constant at 10\% by setting $j = 0.1 p$.}
\label{fig:power.synth.all}
\end{figure}

Figure \ref{fig:power.synth.all} report the results. Panels a, c, and e report the estimated empirical power at $\alpha = 0.05$ and shows that the efficient perm. ED-test (blue) dominates the cross-ED test (red) in terms of power. (The error bands around the curves correspond to one bootstrap standard deviation from the mean based on 200 bootstrap samples.) Panels b, d, and f report the computation times. While the cross-ED test is faster than the efficient permutation approach, the small dynamic ranges on the y-axis of these panels show that the absolute difference in computation time is still fairly small. These experiments illustrate how the efficient perm. ED-test is able to trade small increases in computation time by sizable increases in statistical power when compared to the cross-ED test.

\section{Discussion}

In this paper, we describe a novel computationally efficient permutation test for multivariate two-sample tests based on energy distance or maximum mean discrepancy statistics. While, we only illustrate the application of the test using the energy distance statistic, the proposed approach is also directly applicable to MMD statistics. Comparisons against the cross-ED test (which represents the current SOTA with regard to computation-statistics trade-off) favor our proposed test since it achieves the best possible statistical power at the expense of a small increase in computation.

Since two-sample testing and statistical independence testing can be reduced to each other, a natural extension of the present work would be to adapt the proposed efficient permutation strategy to statistical independence testing. While this can potentially reduce the computation time relative to standard permutation approaches for independence testing, we anticipate that the improvements won't be as considerable as for two-sample testing. To see why, note that the computation of ED or MMD statistics is very fast once the Euclidean distance (kernel) matrices are computed, since they are accomplished by simple averaging operations over the elements of these matrices (as it is clear from equations \ref{eq:ed.sample.estimator} and \ref{eq:mmd.biased.sample.estimator}). The computation of the test statistics for independence testing, such as the HSIC~\cite{Gretton2007} statistic, on the other hand, have two expensive steps: (i) the computation of the kernel matrices $\bfK_{\bfX \bfX}$ and $\bfK_{\bfY \bfY}$; and (ii) the computation of the statistic per se,
$$
\mbox{HSIC} = \frac{1}{n-3} \left[ tr(\tilde{\bfK}_{\bfX \bfX} \tilde{\bfK}_{\bfY \bfY}) + \frac{\bfind^T \tilde{\bfK}_{\bfX \bfX} \bfind \bfind^T \tilde{\bfK}_{\bfY \bfY} \bfind}{(n - 1)(n - 2)} - \frac{1}{n - 2} \bfind^T \tilde{\bfK}_{\bfX \bfX} \tilde{\bfK}_{\bfY \bfY} \bfind \right]~,
$$
where $\tilde{\bfK}_{\bfX \bfX} = \bfK_{\bfX \bfX} - diag(\bfK_{\bfX \bfX})$ (and similarly for $\tilde{\bfK}_{\bfY \bfY}$), $n = n_x = n_y$, and $\bfind$ represents a vector of 1s of dimension $n \times 1$. Both of these steps have quadratic time complexity on the sample size~\cite{Song2012}. Hence, while we can easily speedup the computation of step (i), the computation of step (ii) remains costly.

\appendix

\section{Appendix}

\subsection{Time complexity of the Euclidean distance matrix}

Consider Algorithm \ref{alg:num.euclid.dist} for the computation of the matrix of Euclidean distances between two multivariate datasets $\bfx$ with dimension $n_x \times p$ and $\bfy$ with dimension $n_y \times p$. For each pair of multidimensional datapoints $\bfx_i$ and $\bfy_j$, the computation of the Euclidean distance between two vectors with $p$ elements (line 5) involves $3 p$ operations, namely, $p$ subtractions, $p$ multiplications (squaring operations), $p - 1$ additions, and 1 square-root computation. Hence, line 5 has complexity $O(p)$. Since the algorithm computes Euclidean distances across all pairwise combinations of indexes $i = 1, \ldots, n_x$ and $j = 1, \ldots, n_y$ (lines 1 to 5) it follows that the algorithm's total complexity is of order $O(n_x n_y p)$. This implies that for fixed number of columns ($p$) the complexity scales quadratically with sample sizes, while for fixed number of rows ($n_x$ and $n_y$) the complexity scales linearly with the number of variables.
\begin{algorithm}
   \caption{$EuclideanDistanceMatrix(\bfx, \bfy)$}
   \label{alg:num.euclid.dist}
   \KwInput{Datasets $\bfx$ and $\bfy$ of dimension $n_x \times p$ and $n_y \times p$, respectively}
   \For{$i=1$ to $n_x$} {
     $\bfx_i \leftarrow \bfx[i,]$\;
     \For{$j=1$ to $n_y$} {
       $\bfy_j \leftarrow \bfy[j,]$\;
       $D[i, j] \leftarrow (\sum_{k = 1}^{p} (x_{ik} - y_{jk})^2)^{1/2}$\;
     }
   }
   \KwOutput{euclidean distance matrix, $\bfD_{\bfX \bfY}$}
\end{algorithm}

Similarly, the computation of Gaussian kernel matrices has the same time complexity (since the computation of the Gaussian kernel between two vectors with $p$ elements also has $O(p)$ complexity, as it involves the same operations for the computation of the Euclidean distance with one additional division by the bandwidth parameter and the replacement of the square-root operation by taking the exponential of the negative scaled Euclidean distance).

\subsection{Standard and pre-computed permutation test algorithms for the ED statistic}

Algorithm \ref{alg:1} describes the standard permutation multivariate two-sample test based on the energy distance statistic. Note that the expensive computations of the Euclidean distance matrices are repeated $b + 1$ times (once for the original data in lines 1 to 3, and $b$ times for the permuted datasets in lines 15 to 17.)

\begin{algorithm}[!h]
\caption{Permutation two-sample test for the energy distance statistic (standard approach)}\label{alg:1}
\KwData{Datasets $\bfx$ and $\bfy$; and number of permutations, $b$}
\tcc{Compute the Euclidean distance matrices on the original datasets.}
\ShowLn {$\bfD_{\bfX \bfX} \leftarrow \mbox{EuclideanDistanceMatrix}(\bfx, \bfx)$} \\
\ShowLn {$\bfD_{\bfY \bfY} \leftarrow \mbox{EuclideanDistanceMatrix}(\bfy, \bfy)$} \\
\ShowLn {$\bfD_{\bfX \bfY} \leftarrow \mbox{EuclideanDistanceMatrix}(\bfx, \bfy)$} \\
\tcc{Compute the ED statistic on the original data.}
\ShowLn {$n_x \leftarrow \mbox{NumberOfRows}(\bfx)$ \tcp{Find number of samples of $\bfx$.}}
\ShowLn {$n_y \leftarrow \mbox{NumberOfRows}(\bfy)$ \tcp{Find number of samples of $\bfy$.}}
\ShowLn {$\mbox{ED}_{o} \leftarrow \frac{2}{n_x n_x} \sum_{k=1}^{n_x} \sum_{l=1}^{n_y} \bfD_{\bfX \bfY}^{kl} - \frac{1}{n_x^2} \sum_{k=1}^{n_x} \sum_{l=1}^{n_x} \bfD_{\bfX \bfX}^{kl} - \frac{1}{n_y^2} \sum_{k=1}^{n_y} \sum_{l=1}^{n_y} \bfD_{\bfY \bfY}^{kl}$} \\
\tcc{Concatenate $\bfx$ and $\bfy$ into a new dataset $\bfw$.}
\ShowLn {$\bfw = (\bfx, \bfy)^T$} \\
\tcc{Create vector of indexes of the concatenated data.}
\ShowLn {$\bfi_w \leftarrow (1, \ldots, n)$, where $n = n_x + n_y$} \\
\ShowLn {$\mbox{ED}^\ast \leftarrow [\;]$ \tcp{Create empty vector.}}
\ShowLn \For{$i$ in 1, \ldots, b} {
  \ShowLn {$\bfai_{x} \leftarrow \mbox{SampleWithoutReplacement}(\bfi_w, n_x)$ \tcp{Randomly sample $n_x$ elements from the $\bfi_w$ vector. This corresponds to the samples of the concatenated data that are assigned to $\bfax$ during the permutation process.}}
  \ShowLn {$\bfai_{y} \leftarrow \mbox{SetDifference}(\bfi_w, \bfi_{x})$ \tcp{Get remaining indexes by taking the set difference between $\bfi_w$ and $\bfai_{x}$. This corresponds to the samples of the concatenated data that are assigned to $\bfay$ during the permutation process.}}
  \ShowLn {$\bfx^\ast \leftarrow \bfc[\bfai_{x},]$ \tcp{Assign the $\bfai_x$ rows of $\bfw$ to $\bfax$.}}
  \ShowLn {$\bfy^\ast \leftarrow \bfc[\bfai_{y},]$ \tcp{Assign the $\bfai_y$ rows of $\bfw$ to $\bfay$.}}
  \tcc{Recompute the Euclidean distance matrices on the permuted data.}
  \ShowLn {$\bfD_{\bfaX \bfaX} \leftarrow \mbox{EuclideanDistanceMatrix}(\bfx^\ast, \bfx^\ast)$} \\
  \ShowLn {$\bfD_{\bfaY \bfaY} \leftarrow \mbox{EuclideanDistanceMatrix}(\bfy^\ast, \bfy^\ast)$} \\
  \ShowLn {$\bfD_{\bfaX \bfaY} \leftarrow \mbox{EuclideanDistanceMatrix}(\bfx^\ast, \bfy^\ast)$} \\
  \tcc{Compute the ED statistic on the permuted data.}
  \ShowLn {$\mbox{ED}^\ast[i] \leftarrow \frac{2}{n_x n_y} \sum_{k=1}^{n_x} \sum_{l=1}^{n_y} \bfD_{\bfaX \bfaY}^{kl} - \frac{1}{n_x^2} \sum_{k=1}^{n_x} \sum_{l=1}^{n_x} \bfD_{\bfaX \bfaX}^{kl} - \frac{1}{n_y^2} \sum_{k=1}^{n_y} \sum_{l=1}^{n_y} \bfD_{\bfaY \bfaY}^{kl}$} \\
}
\tcc{Compute the permutation p-value.}
\ShowLn {$\mbox{perm. p-value} \leftarrow (1 + \sum_{i=1}^{b} \ind \{ \mbox{ED}^\ast[i] \ge \mbox{ED}_o \})/(b + 1)$} \\
\KwResult{Return the permutation p-value, and permutation null distribution (ED$^\ast$)}
\end{algorithm}

Algorithm \ref{alg:pre} describes the pre-computation strategy for reducing the computational cost of the standard permutation test. The basic idea is to pre-compute the Euclidean distance matrix using the concatenated data, $\bfw = (\bfx, \bfy)^T$, as described in line 3, and then, at each permutation iteration, compute the test statistic on permuted data by simply permuting the indexes of the concatenated data (lines 14 and 15) and extracting the respective elements from the pre-computed matrix (lines 16 to 18), rather than recomputing the matrices on permuted data.

\begin{algorithm}[!h]
\caption{Permutation two-sample test for the energy distance statistic (pre-computation strategy)}\label{alg:pre}
\KwData{Datasets $\bfx$ and $\bfy$; and number of permutations, $b$}
\tcc{Concatenate $\bfx$ and $\bfy$ into a new dataset $\bfw$.}
\ShowLn {$\bfw = (\bfx, \bfy)^T$} \\
\tcc{Create vector containing the indexes of the concatenated data.}
\ShowLn {$\bfi_w \leftarrow (1, \ldots, n_x + n_y)$} \\
\tcc{Compute the Euclidean distance matrix of the concatenated data. Note that this computation is performed only once.}
\ShowLn {$\bfD_{\bfW \bfW} \leftarrow \mbox{EuclideanDistanceMatrix}(\bfw, \bfw)$} \\
\tcc{Compute the ED statistic on the original data.}
\ShowLn {$n_x \leftarrow \mbox{NumberOfRows}(\bfx)$} \\
\ShowLn {$n_y \leftarrow \mbox{NumberOfRows}(\bfy)$} \\
\ShowLn {$\bfi_x \leftarrow (1, \ldots, n_{x})$ \tcp{Indexes of the samples from $\bfx$ (relative to $\bfw$)}}
\ShowLn {$\bfi_y \leftarrow (n_{x} + 1, \ldots, n_x + n_y)$ \tcp{Indexes of the samples from $\bfy$ (relative to $\bfw$)}}
\ShowLn {$\bfD_{\bfX \bfY} \leftarrow \bfD_{\bfW \bfW}[\, \bfi_x \, , \, \bfi_y \,]$ \tcp{Extract $\bfD_{\bfX \bfY}$ from $\bfD_{\bfW \bfW}$.}}
\ShowLn {$\bfD_{\bfX \bfX} \leftarrow \bfD_{\bfW \bfW}[\, \bfi_x \, , \, \bfi_x \,]$ \tcp{Extract $\bfD_{\bfX \bfX}$ from $\bfD_{\bfW \bfW}$.}}
\ShowLn {$\bfD_{\bfY \bfY} \leftarrow \bfD_{\bfW \bfW}[\, \bfi_y \, , \, \bfi_y \,]$ \tcp{Extract $\bfD_{\bfY \bfY}$ from $\bfD_{\bfW \bfW}$.}}
\ShowLn {$\mbox{ED}_{o} \leftarrow \frac{2}{n_x n_x} \sum_{k=1}^{n_x} \sum_{l=1}^{n_y} \bfD_{\bfX \bfY}^{kl} - \frac{1}{n_x^2} \sum_{k=1}^{n_x} \sum_{l=1}^{n_x} \bfD_{\bfX \bfX}^{kl} - \frac{1}{n_y^2} \sum_{k=1}^{n_y} \sum_{l=1}^{n_y} \bfD_{\bfY \bfY}^{kl}$ \tcp{ED statistic on original data.}}
\ShowLn {$\mbox{ED}^\ast \leftarrow [\;]$ \tcp{Create empty vector.}}
\ShowLn \For{$i$ in 1, \ldots, b} {
  \ShowLn {$\bfai_x \leftarrow \mbox{SampleWithoutReplacement}(\bfi_w, n_x)$ \tcp{Randomly sample $n_x$ elements from the $\bfi_w$ vector. This corresponds to the samples of the concatenated data that are assigned to $\bfax$ during the permutation process.}}
  \ShowLn {$\bfai_y \leftarrow \mbox{SetDifference}(\bfi_w, \bfai_x)$ \tcp{Get remaining indexes. This corresponds to the samples of the concatenated data that are assigned to $\bfay$ during the permutation process.}}
  \ShowLn {$\bfD_{\bfaX \bfaY} \leftarrow \bfD_{\bfW \bfW}[\, \bfai_x \, , \, \bfai_y \,]$ \tcp{Extract $\bfD_{\bfaX \bfaY}$ from $\bfD_{\bfW \bfW}$.}}
  \ShowLn {$\bfD_{\bfaX \bfaX} \leftarrow \bfD_{\bfW \bfW}[\, \bfai_x \, , \, \bfai_x \,]$ \tcp{Extract $\bfD_{\bfaX \bfaX}$ from $\bfD_{\bfW \bfW}$.}}
  \ShowLn {$\bfD_{\bfaY \bfaY} \leftarrow \bfD_{\bfW \bfW}[\, \bfai_y \, , \, \bfai_y \,]$ \tcp{Extract $\bfD_{\bfaY \bfaY}$ from $\bfD_{\bfW \bfW}$.}}
  \ShowLn {$\mbox{ED}^\ast[i] \leftarrow \frac{2}{n_x n_y} \sum_{k=1}^{n_x} \sum_{l=1}^{n_y} \bfD_{\bfaX \bfaY}^{kl} - \frac{1}{n_x^2} \sum_{k=1}^{n_x} \sum_{l=1}^{n_x} \bfD_{\bfaX \bfaX}^{kl} - \frac{1}{n_y^2} \sum_{k=1}^{n_y} \sum_{l=1}^{n_y} \bfD_{\bfaY \bfaY}^{kl}$ \tcp{ED statistic on the permuted data.}}
}
\tcc{Compute the permutation p-value.}
\ShowLn {$\mbox{perm. p-value} \leftarrow (1 + \sum_{i=1}^{b} \ind \{ \mbox{ED}^\ast[i] \ge \mbox{ED}_o \})/(b + 1)$} \\
\KwResult{Return the permutation p-value, and permutation null distribution (ED$^\ast$)}
\end{algorithm}

\subsection{The cross-ED test}

As described in reference~\cite{Shekhar2022}, the two-sample cross kernel MMD statistic, $xMMD^2$, is computed as,
\begin{equation}
xMMD^2 = \frac{1}{n_{x_1} \, n_{x_2} \, n_{y_1} \, n_{y_2}} \sum_{i=1}^{n_{x_1}} \sum_{i'=1}^{n_{x_2}} \sum_{j=1}^{n_{y_1}} \sum_{j'=1}^{n_{x_2}} h_K(\bfx_i, \bfx_{i'}, \bfy_j, \bfy_{j'})~,
\end{equation}
with the kernel function $h_K(.)$ given by,
\begin{align*}
h_K&(\bfx_i, \bfx_{i'}, \bfy_j, \bfy_{j'}) \\
&= K(\bfx_i, \bfx_{i'}) - K(\bfx_i, \bfy_{j'}) - K(\bfy_j, \bfx_{i'}) + K(\bfy_j, \bfy_{j'})~.
\end{align*}
The final test statistic, $\bar{x}MMD^2 = xMMD^2/\hat{\sigma}$, corresponds to a studentized version of $xMMD^2$, where $\hat{\sigma}$ is an empirical variance estimate (see~\cite{Shekhar2022} for details).

A computationally efficient algorithm for the computation of the $\bar{x}MMD^2$ test is given by Algorithm \ref{alg:xmmd}\footnote{A Python implementation of this algorithm (accompanying reference~\cite{Shekhar2022}) is provided in \texttt{https://github.com/sshekhar17/PermFreeMMD/blob/main/src/utils.py}}.
\begin{algorithm}[!h]
\caption{Cross-MMD test}\label{alg:xmmd}
\KwData{Datasets $\bfx$ and $\bfy$}
\ShowLn $n_x \leftarrow \mbox{NumberOfRows}(\bfx)$ \\
\ShowLn $n_y \leftarrow \mbox{NumberOfRows}(\bfy)$ \\
\ShowLn $n_{x_1} \leftarrow \mbox{Floor}(n_x / 2)$ \\
\ShowLn $n_{y_1} \leftarrow \mbox{Floor}(n_y / 2)$ \\
\ShowLn $\bfx_1 \leftarrow \bfx[1:n_{x_1},] $ \\
\ShowLn $\bfx_2 \leftarrow \bfx[(n_{x_1} + 1):n_x,] $ \\
\ShowLn $\bfy_1 \leftarrow \bfy[1:n_{y_1},] $ \\
\ShowLn $\bfy_2 \leftarrow \bfy[(n_{y_1} + 1):n_y,] $ \\
\ShowLn $\bfK_{\bfX_1 \bfX_2} \leftarrow \mbox{KernelFunction}(\bfx_1, \bfx_2)$ \\
\ShowLn $\bfK_{\bfY_1 \bfY_2} \leftarrow \mbox{KernelFunction}(\bfy_1, \bfy_2)$ \\
\ShowLn $\bfK_{\bfX_1 \bfY_2} \leftarrow \mbox{KernelFunction}(\bfx_1, \bfy_2)$ \\
\ShowLn $\bfK_{\bfY_1 \bfX_2} \leftarrow \mbox{KernelFunction}(\bfy_1, \bfx_2)$ \\
\ShowLn $U_x \leftarrow \mbox{Mean}(\bfK_{\bfX_1 \bfX_2}) - \mbox{Mean}(\bfK_{\bfX_1 \bfY_2})$ \\
\ShowLn $U_y \leftarrow \mbox{Mean}(\bfK_{\bfY_1 \bfX_2}) - \mbox{Mean}(\bfK_{\bfY_1 \bfY_2})$ \\
\ShowLn $\hat{U} \leftarrow U_x - U_y$ \\
\ShowLn $\hat{\sigma}_x \leftarrow \mbox{Mean}((\mbox{RowMeans}(\bfK_{\bfX_1 \bfX_2}) - \mbox{RowMeans}(\bfK_{\bfX_1 \bfY_2}) - U_x)^2)$ \\
\ShowLn $\hat{\sigma}_y \leftarrow \mbox{Mean}((\mbox{RowMeans}(\bfK_{\bfY_1 \bfX_2}) - \mbox{RowMeans}(\bfK_{\bfY_1 \bfY_2}) - U_y)^2)$ \\
\ShowLn $\hat{\sigma} \leftarrow \sqrt{\hat{\sigma}_x/n_{x_1} + \hat{\sigma}_y/n_{y_1}}$ \\
\ShowLn $\mbox{p-value} \leftarrow 1 - \Phi(\hat{U}/\hat{\sigma})$ \\
\KwResult{Return the p-value}
\end{algorithm}

Now, as described in~\cite{SzekelyRizzo2004}, the energy distance statistic can be expressed in V-statistic form as,
\begin{equation}
xED = \frac{1}{n_{x_1} \, n_{x_2} \, n_{y_1} \, n_{y_2}} \sum_{i=1}^{n_{x_1}} \sum_{i'=1}^{n_{x_2}} \sum_{j=1}^{n_{y_1}} \sum_{j'=1}^{n_{x_2}} h_D(\bfx_i, \bfx_{i'}, \bfy_j, \bfy_{j'})~,
\label{eq:xed.h}
\end{equation}
with the kernel function $h_D(.)$ given by,
\begin{align*}
h_D&(\bfx_i, \bfx_{i'}, \bfy_j, \bfy_{j'}) \\
&= D(\bfx_i, \bfy_{j'}) - D(\bfx_i, \bfx_{i'}) - D(\bfy_j, \bfy_{j'}) + D(\bfy_j, \bfx_{i'})~,
\end{align*}
it follows that the corresponding efficient algorithm for the computation of the cross kernel test based on the energy distance statistic (denoted cross-ED test) is given by Algorithm \ref{alg:xed}.
\begin{algorithm}[!h]
\caption{Cross-ED test}\label{alg:xed}
\KwData{Datasets $\bfx$ and $\bfy$}
\ShowLn $n_x \leftarrow \mbox{NumberOfRows}(\bfx)$ \\
\ShowLn $n_y \leftarrow \mbox{NumberOfRows}(\bfy)$ \\
\ShowLn $n_{x_1} \leftarrow \mbox{Floor}(n_x / 2)$ \\
\ShowLn $n_{y_1} \leftarrow \mbox{Floor}(n_y / 2)$ \\
\ShowLn $\bfx_1 \leftarrow \bfx[1:n_{x_1},] $ \\
\ShowLn $\bfx_2 \leftarrow \bfx[(n_{x_1} + 1):n_x,] $ \\
\ShowLn $\bfy_1 \leftarrow \bfy[1:n_{y_1},] $ \\
\ShowLn $\bfy_2 \leftarrow \bfy[(n_{y_1} + 1):n_y,] $ \\
\ShowLn $\bfD_{\bfX_1 \bfX_2} \leftarrow \mbox{EuclideanDistanceMatrix}(\bfx_1, \bfx_2)$ \\
\ShowLn $\bfD_{\bfY_1 \bfY_2} \leftarrow \mbox{EuclideanDistanceMatrix}(\bfy_1, \bfy_2)$ \\
\ShowLn $\bfD_{\bfX_1 \bfY_2} \leftarrow \mbox{EuclideanDistanceMatrix}(\bfx_1, \bfy_2)$ \\
\ShowLn $\bfD_{\bfY_1 \bfX_2} \leftarrow \mbox{EuclideanDistanceMatrix}(\bfy_1, \bfx_2)$ \\
\ShowLn $U_x \leftarrow \mbox{Mean}(\bfD_{\bfX_1 \bfY_2}) - \mbox{Mean}(\bfD_{\bfX_1 \bfX_2})$ \\
\ShowLn $U_y \leftarrow \mbox{Mean}(\bfD_{\bfY_1 \bfY_2}) - \mbox{Mean}(\bfD_{\bfY_1 \bfX_2})$ \\
\ShowLn {$\hat{U} \leftarrow U_x - U_y$ \tcp{$\hat{U}$ corresponds to statistic in equation \ref{eq:xed.h}}}
\ShowLn $\hat{\sigma}_x \leftarrow \mbox{Mean}((\mbox{RowMeans}(\bfD_{\bfX_1 \bfY_2}) - \mbox{RowMeans}(\bfD_{\bfX_1 \bfX_2}) - U_x)^2)$ \\
\ShowLn $\hat{\sigma}_y \leftarrow \mbox{Mean}((\mbox{RowMeans}(\bfD_{\bfY_1 \bfY_2}) - \mbox{RowMeans}(\bfD_{\bfY_1 \bfX_2}) - U_y)^2)$ \\
\ShowLn $\hat{\sigma} \leftarrow \sqrt{\hat{\sigma}_x/n_{x_1} + \hat{\sigma}_y/n_{y_1}}$ \\
\ShowLn {$\mbox{p-value} \leftarrow 1 - \Phi(\hat{U}/\hat{\sigma})$ \tcp{$\Phi()$ corresponds to the cumulative distribution function of a $N(0, 1)$ random variable.}}
\KwResult{Return the p-value}
\end{algorithm}

Note, as well, that an alternative way to compute the cross-ED test is to simply change the kernel matrices $\bfK_{\bfX_1 \bfX_2}$, $\bfK_{\bfY_1 \bfY_2}$, $\bfK_{\bfX_1 \bfY_2}$, and $\bfK_{\bfY_1 \bfX_2}$ in Algorithm \ref{alg:xmmd} by the respective Euclidean distance matrices (namely, $\bfD_{\bfX_1 \bfX_2}$, $\bfD_{\bfY_1 \bfY_2}$, $\bfD_{\bfX_1 \bfY_2}$, and $\bfD_{\bfY_1 \bfX_2}$), and change line 19 of Algorithm \ref{alg:xmmd} to $1 - \Phi(-\hat{U}/\hat{\sigma})$.

\subsection{Illustrating the application of Algorithms 1 and 2}

Here, we provide a step-by-step illustration of the application of Algorithms \ref{alg:eff} and \ref{alg:eff2} to the toy dataset example in in equation (\ref{eq:original.data}) and its permuted version in equation (\ref{eq:permuted.data}).

Lines 1 to 3 of Algorithm \ref{alg:eff} show that the Euclidean distance matrices are computed only once using the original data (producing the matrices in equations \ref{eq:euclid.matrix.x.x}, \ref{eq:euclid.matrix.y.y}, and \ref{eq:euclid.matrix.x.y}), while line 4 shows that the matrix $\bfD_{\bfY \bfX}$ is obtained by transposing matrix $\bfD_{\bfX \bfY}$.

Lines 5, 6, and 7 of Algorithm \ref{alg:eff} describe the computation of the energy distance statistic on the original data.

Lines 8 to 26 of Algorithm \ref{alg:eff} describe the computation of the permutation null distribution. Line 8 creates an empty vector for storing the ED statistics computed on each of the $b$ data permutations. At each iteration $i$ of the for-loop, line 10 calls the function \texttt{PermutationIndexesTwoSampleTest}, implemented in Algorithm \ref{alg:eff2}, which finds the mapping between the entries of the Euclidean distance matrices computed on the original data and the Euclidean distance matrices we would have obtained using the permuted data, allowing us to reconstruct the permuted data Euclidean distance matrices by swapping entries without having to directly compute these matrices.

Continuing with our toy example, we have from lines 2 to 5 of Algorithm \ref{alg:eff2} that,
$$
idx_w = \left(\hspace{-0.1cm}
\begin{array}{c}
1 \\
2 \\
3 \\
4 \\
5 \\
6 \\
7 \\
8 \\
9 \\
\end{array}
\hspace{-0.1cm} \right),
\hspace{0.3cm}
idx_g = \left(\hspace{-0.1cm}
\begin{array}{c}
1 \\
2 \\
3 \\
4 \\
5 \\
1 \\
2 \\
3 \\
4 \\
\end{array}
\hspace{-0.1cm} \right),
\hspace{0.3cm}
idx_{p1} = \left(\hspace{-0.1cm}
\begin{array}{c}
7 \\
4 \\
5 \\
6 \\
2 \\
\end{array}
\hspace{-0.1cm} \right),
\hspace{0.3cm}
idx_{p2} = \left(\hspace{-0.1cm}
\begin{array}{c}
1 \\
3 \\
8 \\
9 \\
\end{array}
\hspace{-0.1cm} \right).
$$
Note that: $idx_w$ corresponds to the row indexes of the concatenated data $\bfw$; the vector $idx_{p1}$ contains the indexes of the datapoints (relative to $\bfw$) which were assigned to $\bfax$ during the permutation process; and $idx_{p2}$ contains the indexes of the datapoints assigned to $\bfay$. Note that the particular values of the $idx_{p1}$ and $idx_{p2}$ above represent the indexes of the data permutation which generated the permuted datasets $\bfax$ and $\bfay$ in equation \ref{eq:permuted.data}, i.e.,
$$
\bfw = \left(\hspace{-0.1cm}
\begin{array}{c}
\bfx_1 \\
\bfx_2 \\
\bfx_3 \\
\bfx_4 \\
\bfx_5 \\
\bfy_1 \\
\bfy_2 \\
\bfy_3 \\
\bfy_4 \\
\end{array}
\hspace{-0.1cm} \right),
\hspace{0.3cm}
\bfw[idx_{p1},] = \left(\hspace{-0.1cm}
\begin{array}{c}
\bfy_2 \\
\bfx_4 \\
\bfx_5 \\
\bfy_1 \\
\bfx_2 \\
\end{array}
\hspace{-0.1cm} \right),
\hspace{0.3cm}
\bfw[idx_{p2},] = \left(\hspace{-0.1cm}
\begin{array}{c}
\bfx_1 \\
\bfx_3 \\
\bfy_3 \\
\bfy_4 \\
\end{array}
\hspace{-0.1cm} \right).
$$

However, in order to reconstruct the Euclidean distance matrices associated with the permuted data using the Euclidean distance values already computed on the original data, it is necessary to map the indexes of $idx_{p1}$ and $idx_{p2}$ to the indexes of samples from $\bfx$ and $\bfy$. (Note that it suffices to map $idx_{p1}$ and $idx_{p2}$ to the row indexes of $\bfx$ and $\bfy$ to get the mapping to the distance matrices since the row indexes of $\bfx$ correspond to the column indexes of $\bfD_{\bfY \bfX}$, row indexes of $\bfD_{\bfX \bfY}$, and row and column indexes of $\bfD_{\bfX \bfX}$, while the row indexes of $\bfy$ correspond to the column indexes of $\bfD_{\bfX \bfY}$, row indexes of $\bfD_{\bfY \bfX}$, and row and column indexes of $\bfD_{\bfY \bfY}$.) Lines 6 and 8 of Algorithm \ref{alg:eff2} describe, respectively, how to obtain the indexes of the samples from the $\bfx$ and $\bfy$ datasets, which were assigned to $\bfax$ during the permutation process. Similarly lines 10 and 12 explain how to obtain the indexes from samples from $\bfx$ and $\bfy$ datasets, which were assigned to $\bfay$.

To illustrate, we now continue with our example. Note that because the concatenated data $\bfw$ is obtained by appending $\bfy$ to the bottom of $\bfx$, we have that any element of $idx_{p1}$ (or $idx_{p2}$) which is less or equal to $n_x = 5$ corresponds to a sample from $\bfx$, whereas any elements greater than 5 correspond to a sample from $\bfy$.

Hence, starting with $\bfax$, we have from line 6 that the positions of $idx_{p1}$ which correspond to samples from the $\bfx$ dataset are given by $\mbox{Which}(idx_{p1} \le 5) = [2, 3, 5]$. Hence, the indexes of the samples from $\bfx$ which were assigned to $\bfax$ relative to the concatenated data $\bfw$ (indexed by $idx_w$) are given by $idx_{p1}[2, 3, 5] = [4, 5, 2]$. Now, because the Euclidean distance matrices are indexed from 1 to $n_x$ (or to $n_y$), rather than from 1 to $n$ (as the concatenated data), we still need to map these indexes back to the $idx_g$ vector. Hence, the indexes of the samples from $\bfx$ which were assigned to $\bfax$ (relative to the indexes of $\bfx$) are given by $\bfi_1 = idx_g[4, 5, 2] = [4, 5, 2]$.

Line 8 describe how to obtain the indexes of the samples from $\bfy$ which were assigned to $\bfax$. Now, the positions of $idx_{p1}$ which correspond to samples from $\bfy$ are given by $\mbox{Which}(idx_{p1} > 5) = [1, 4]$. Hence, the indexes of the samples from $\bfy$ which were assigned to $\bfax$ (relative to the concatenated data) are given by $idx_{p1}[1, 4] = [7, 6]$. Now, mapping back to the indexes of $\bfy$, we have that $\bfi_2 = idx_g[7, 6] = [2, 1]$.

Since the lengths of $\bfi_1$ and $\bfi_2$ are given by 3 and 2, respectively, we have that the corresponding indexes on the reconstructed permuted data Euclidean distance matrices are given, respectively, by  $\bfi_1^\ast = [1, 2, 3]$ and $\bfi_2^\ast = [4, 5]$, as described in lines 7 and 9 of Algorithm \ref{alg:eff2}.

Lines 10 to 13 of Algorithm \ref{alg:eff2} describe the analogous operations to obtain the indexes from samples of the $\bfx$ and $\bfy$ datasets which were assigned to $\bfay$ during the permutation process. Briefly, from line 10 we have that the positions at which $idx_{p2} \le 5$ are 1 and 2. Hence, $idx_{p2}[1, 2] = [1, 3]$ and $\bfj_1 = idx_g[1, 3] = [1, 3]$. Since the length of $\bfj_1$ is 2, we have from line 11 that the corresponding indexes in the permuted data distance matrices are $\bfj_1^\ast = [1, 2]$. From line 12 we have that the positions at which $idx_{p2} > 5$ are 3 and 4. Hence, $idx_{p2}[3, 4] = [8, 9]$ and $\bfj_2 = idx_g[8, 9] = [3, 4]$. Since the length of $\bfj_2$ is 2, we have from line 13 that the corresponding indexes in the permuted data distance matrices are $\bfj_2^\ast = [3, 4]$.

In summary, application of Algorithm \ref{alg:eff2} to the toy permutation example described in equations \ref{eq:original.data} and \ref{eq:permuted.data}, generates the following sets of indexes,
$$
\begin{array}{l}
\bfi_1 = [4, 5, 2], \\
\bfi_2 = [2, 1], \\
\bfai_1 = [1, 2, 3], \\
\bfai_1 = [4, 5], \\
\end{array}
\hspace{1.0cm}
\begin{array}{l}
\bfj_1 = [1, 3], \\
\bfj_2 = [3, 4], \\
\bfaj_1 = [1, 2], \\
\bfaj_1 = [3, 4], \\
\end{array}
$$
where the permuted samples assigned to $\bfax$ are given by,
$$
\bfx[\bfi_1,] = \left(\hspace{-0.1cm}
\begin{array}{c}
\bfx_4 \\
\bfx_5 \\
\bfx_2 \\
\end{array}
\hspace{-0.1cm} \right),
\hspace{0.3cm}
\bfy[\bfi_2,] = \left(\hspace{-0.1cm}
\begin{array}{c}
\bfy_2 \\
\bfy_1 \\
\end{array}
\hspace{-0.1cm} \right),
$$
so that,
\begin{equation}
\bfx^\ast = \left(\hspace{-0.1cm}
\begin{array}{c}
\bfx[\bfi_1,] \\
\bfy[\bfi_2,] \\
\end{array}
\hspace{-0.1cm} \right) =
\left(\hspace{-0.1cm}
\begin{array}{c}
\bfx_4 \\
\bfx_5 \\
\bfx_2 \\
\bfy_2 \\
\bfy_1 \\
\end{array}
\hspace{-0.1cm} \right),
\label{eq:permuted.data.2}
\end{equation}
while the permuted samples assigned to $\bfay$ are given by,
$$
\bfx[\bfj_1,] = \left(\hspace{-0.1cm}
\begin{array}{c}
\bfx_1 \\
\bfx_3 \\
\end{array}
\hspace{-0.1cm} \right),
\hspace{0.3cm}
\bfy[\bfj_2,] = \left(\hspace{-0.1cm}
\begin{array}{c}
\bfy_3 \\
\bfy_4 \\
\end{array}
\hspace{-0.1cm} \right),
$$
so that,
\begin{equation}
\bfy^\ast = \left(\hspace{-0.1cm}
\begin{array}{c}
\bfx[\bfj_1,] \\
\bfy[\bfj_2,] \\
\end{array}
\hspace{-0.1cm} \right) =
\left(\hspace{-0.1cm}
\begin{array}{c}
\bfx_1 \\
\bfx_3 \\
\bfy_3 \\
\bfy_4 \\
\end{array}
\hspace{-0.1cm} \right).
\end{equation}
Note that while the order of the row elements of $\bfax$ in equation \ref{eq:permuted.data.2} (generated by Algorithm \ref{alg:eff2}) is different from the order of the elements of $\bfax$ in equation \ref{eq:permuted.data}, the elements are still the same.

Now, going back to Algorithm \ref{alg:eff}, we have from lines 11 to 15 that the reconstructed Euclidean distance matrix $\bfD_{\bfX^\ast \bfX^\ast}$ is given by,
\begin{align*}
\bfD_{\bfX^\ast \bfX^\ast} &= \left(
\arraycolsep=3.0pt\def\arraystretch{2.0}
\begin{array}{c|c}
\bfD_{\bfaX \bfaX}[\bfai_1, \bfai_1] & \bfD_{\bfaX \bfaX}[\bfai_1, \bfai_2] \\ \hline
\bfD_{\bfaX \bfaX}[\bfai_2, \bfai_1] & \bfD_{\bfaX \bfaX}[\bfai_2, \bfai_2] \\
\end{array}
\right) =
\left(
\arraycolsep=3.0pt\def\arraystretch{2.0}
\begin{array}{c|c}
\bfD_{\bfX \bfX}[\bfi_1, \bfi_1] & \bfD_{\bfX \bfY}[\bfi_1, \bfi_2] \\ \hline
\bfD_{\bfY \bfX}[\bfi_2, \bfi_1] & \bfD_{\bfY \bfY}[\bfi_2, \bfi_2] \\
\end{array}
\right),
\end{align*}
where,
\begin{align*}
\bfD_{\bfaX \bfaX}[\bfai_1, \bfai_1] &= \bfD_{\bfX \bfX}[\bfi_1, \bfi_1] =\left(\hspace{-0.2cm}
\begin{array}{ccc}
D(\bfx_4,\bfx_4) & D(\bfx_4,\bfx_5) & D(\bfx_4,\bfx_2) \\
D(\bfx_5,\bfx_4) & D(\bfx_5,\bfx_5) & D(\bfx_5,\bfx_2) \\
D(\bfx_2,\bfx_4) & D(\bfx_2,\bfx_5) & D(\bfx_2,\bfx_2) \\
\end{array}
\hspace{-0.2cm} \right),
\end{align*}
$$
\bfD_{\bfaX \bfaX}[\bfai_1, \bfai_2] = \bfD_{\bfX \bfY}[\bfi_1, \bfi_2] =
\left(\hspace{-0.2cm}
\begin{array}{ccccc}
D(\bfx_4,\bfy_2) & D(\bfx_4,\bfy_1) \\
D(\bfx_5,\bfy_2) & D(\bfx_5,\bfy_1) \\
D(\bfx_2,\bfy_2) & D(\bfx_2,\bfy_1) \\
\end{array}
\hspace{-0.2cm} \right),
$$
\begin{align*}
\bfD_{\bfaX \bfaX}[\bfai_2, \bfai_1] &= \bfD_{\bfY \bfX}[\bfi_2, \bfi_1] = \bfD_{\bfX \bfY}[\bfi_1, \bfi_2]^T = \left(\hspace{-0.2cm}
\begin{array}{ccccc}
D(\bfx_4,\bfy_2) & D(\bfx_5,\bfy_2) & D(\bfx_2,\bfy_2) \\
D(\bfx_4,\bfy_1) & D(\bfx_5,\bfy_1) & D(\bfx_2,\bfy_1) \\
\end{array}
\hspace{-0.2cm} \right),
\end{align*}
$$
\bfD_{\bfaX \bfaX}[\bfai_2, \bfai_2] = \bfD_{\bfY \bfY}[\bfi_2, \bfi_2] =
\left(\hspace{-0.2cm}
\begin{array}{cccc}
D(\bfy_2,\bfy_2) & D(\bfy_2,\bfy_1) \\
D(\bfy_1,\bfy_2) & D(\bfy_1,\bfy_1) \\
\end{array}
\hspace{-0.2cm} \right),
$$
so that,
\begin{align}
&\bfD_{\bfaX \bfaX} = \left(\hspace{-0.2cm}
\begin{array}{ccccc}
D(\bfx_4,\bfx_4) & D(\bfx_4,\bfx_5) & D(\bfx_4,\bfx_2) & D(\bfx_4,\bfy_2) & D(\bfx_4,\bfy_1) \\
D(\bfx_5,\bfx_4) & D(\bfx_5,\bfx_5) & D(\bfx_5,\bfx_2) & D(\bfx_5,\bfy_2) & D(\bfx_5,\bfy_1) \\
D(\bfx_2,\bfx_4) & D(\bfx_2,\bfx_5) & D(\bfx_2,\bfx_2) & D(\bfx_2,\bfy_2) & D(\bfx_2,\bfy_1) \\
D(\bfx_4,\bfy_2) & D(\bfx_5,\bfy_2) & D(\bfx_2,\bfy_2) & D(\bfy_2,\bfy_2) & D(\bfy_2,\bfy_1) \\
D(\bfx_4,\bfy_1) & D(\bfx_5,\bfy_1) & D(\bfx_2,\bfy_1) & D(\bfy_1,\bfy_2) & D(\bfy_1,\bfy_1) \\
\end{array}
\hspace{-0.2cm} \right).
\label{eq:euclid.matrix.xstar.xstar.2}
\end{align}


From lines 16 to 20, we have that the reconstructed Euclidean distance matrix $\bfD_{\bfY^\ast \bfY^\ast}$ is given by,
\begin{align*}
\bfD_{\bfY^\ast \bfY^\ast} &= \left(
\arraycolsep=3.0pt\def\arraystretch{2.0}
\begin{array}{c|c}
\bfD_{\bfaY \bfaY}[\bfaj_1, \bfaj_1] & \bfD_{\bfaY \bfaY}[\bfaj_1, \bfaj_2] \\ \hline
\bfD_{\bfaY \bfaY}[\bfaj_2, \bfaj_1] & \bfD_{\bfaY \bfaY}[\bfaj_2, \bfaj_2] \\
\end{array}
\right) =
\left(
\arraycolsep=3.0pt\def\arraystretch{2.0}
\begin{array}{c|c}
\bfD_{\bfX \bfX}[\bfj_1, \bfj_1] & \bfD_{\bfX \bfY}[\bfj_1, \bfj_2] \\ \hline
\bfD_{\bfY \bfX}[\bfj_2, \bfj_1] & \bfD_{\bfY \bfY}[\bfj_2, \bfj_2] \\
\end{array}
\right),
\end{align*}
where,
$$
\bfD_{\bfaY \bfaY}[\bfaj_1, \bfaj_1] = \bfD_{\bfX \bfX}[\bfj_1, \bfj_1] =
\left(\hspace{-0.2cm}
\begin{array}{ccccc}
D(\bfx_1,\bfx_1) & D(\bfx_1,\bfx_3) \\
D(\bfx_3,\bfx_1) & D(\bfx_3,\bfx_3) \\
\end{array}
\hspace{-0.2cm} \right),
$$
$$
\bfD_{\bfaY \bfaY}[\bfaj_1, \bfaj_2] = \bfD_{\bfX \bfY}[\bfj_1, \bfj_2] =
\left(\hspace{-0.2cm}
\begin{array}{ccccc}
D(\bfx_1,\bfy_3) & D(\bfx_1,\bfy_4) \\
D(\bfx_3,\bfy_3) & D(\bfx_3,\bfy_4) \\
\end{array}
\hspace{-0.2cm} \right),
$$
\begin{align*}
\bfD_{\bfaY \bfaY}[\bfaj_2, \bfaj_1] &= \bfD_{\bfY \bfX}[\bfj_2, \bfj_1] = \bfD_{\bfX \bfY}[\bfj_1, \bfj_2]^T =
\left(\hspace{-0.2cm}
\begin{array}{ccccc}
D(\bfx_1,\bfy_3) & D(\bfx_3,\bfy_3) \\
D(\bfx_1,\bfy_4) & D(\bfx_3,\bfy_4) \\
\end{array}
\hspace{-0.2cm} \right),
\end{align*}
$$
\bfD_{\bfaY \bfaY}[\bfaj_2, \bfaj_2] = \bfD_{\bfY \bfY}[\bfj_2, \bfj_2] =
\left(\hspace{-0.2cm}
\begin{array}{cccc}
D(\bfy_3,\bfy_3) & D(\bfy_3,\bfy_4) \\
D(\bfy_4,\bfy_3) & D(\bfy_4,\bfy_4) \\
\end{array}
\hspace{-0.2cm} \right),
$$
so that,
\begin{equation}
\bfD_{\bfaY \bfaY} = \left(\hspace{-0.2cm}
\begin{array}{cccc}
D(\bfx_1,\bfx_1) & D(\bfx_1,\bfx_3) & D(\bfx_1,\bfy_3) & D(\bfx_1,\bfy_4) \\
D(\bfx_3,\bfx_1) & D(\bfx_3,\bfx_3) & D(\bfx_3,\bfy_3) & D(\bfx_3,\bfy_4) \\
D(\bfy_3,\bfx_1) & D(\bfy_3,\bfx_3) & D(\bfy_3,\bfy_3) & D(\bfy_3,\bfy_4) \\
D(\bfy_4,\bfx_1) & D(\bfy_4,\bfx_3) & D(\bfy_4,\bfy_3) & D(\bfy_4,\bfy_4) \\
\end{array}
\hspace{-0.2cm} \right).
\label{eq:euclid.matrix.ystar.ystar.2}
\end{equation}


From lines 21 to 25, we have that the reconstructed Euclidean distance matrix $\bfD_{\bfX^\ast \bfY^\ast}$ is given by,
\begin{align*}
\bfD_{\bfX^\ast \bfY^\ast} &= \left(
\arraycolsep=3.0pt\def\arraystretch{2.0}
\begin{array}{c|c}
\bfD_{\bfaX \bfaY}[\bfai_1, \bfaj_1] & \bfD_{\bfaX \bfaY}[\bfai_1, \bfaj_2] \\ \hline
\bfD_{\bfaX \bfaY}[\bfai_2, \bfaj_1] & \bfD_{\bfaX \bfaY}[\bfai_2, \bfaj_2] \\
\end{array}
\right) =
\left(
\arraycolsep=3.0pt\def\arraystretch{2.0}
\begin{array}{c|c}
\bfD_{\bfX \bfX}[\bfi_1, \bfj_1] & \bfD_{\bfX \bfY}[\bfi_1, \bfj_2] \\ \hline
\bfD_{\bfY \bfX}[\bfi_2, \bfj_1] & \bfD_{\bfY \bfY}[\bfi_2, \bfj_2] \\
\end{array}
\right),
\end{align*}
where,
$$
\bfD_{\bfaX \bfaY}[\bfai_1, \bfaj_1] = \bfD_{\bfX \bfX}[\bfi_1, \bfj_1] =
\left(\hspace{-0.2cm}
\begin{array}{ccccc}
D(\bfx_4,\bfx_1) & D(\bfx_4,\bfx_3) \\
D(\bfx_5,\bfx_1) & D(\bfx_5,\bfx_3) \\
D(\bfx_2,\bfx_1) & D(\bfx_2,\bfx_3) \\
\end{array}
\hspace{-0.2cm} \right),
$$
$$
\bfD_{\bfaX \bfaY}[\bfai_1, \bfaj_2] = \bfD_{\bfX \bfY}[\bfi_1, \bfj_2] =
\left(\hspace{-0.2cm}
\begin{array}{ccccc}
D(\bfx_4,\bfy_3) & D(\bfx_4,\bfy_4) \\
D(\bfx_5,\bfy_3) & D(\bfx_5,\bfy_4) \\
D(\bfx_2,\bfy_3) & D(\bfx_2,\bfy_4) \\
\end{array}
\hspace{-0.2cm} \right),
$$
\begin{align*}
\bfD_{\bfaX \bfaY}[\bfai_2, \bfaj_1] &= \bfD_{\bfY \bfX}[\bfi_2, \bfj_1] = \bfD_{\bfX \bfY}[\bfj_1, \bfi_2]^T = \left(\hspace{-0.2cm}
\begin{array}{ccccc}
D(\bfx_1,\bfy_2) & D(\bfx_3,\bfy_2) \\
D(\bfx_1,\bfy_1) & D(\bfx_3,\bfy_1) \\
\end{array}
\hspace{-0.2cm} \right),
\end{align*}
$$
\bfD_{\bfaX \bfaY}[\bfai_2, \bfaj_2] = \bfD_{\bfY \bfY}[\bfi_2, \bfj_2] =
\left(\hspace{-0.2cm}
\begin{array}{cccc}
D(\bfy_2,\bfy_3) & D(\bfy_2,\bfy_4) \\
D(\bfy_1,\bfy_3) & D(\bfy_1,\bfy_4) \\
\end{array}
\hspace{-0.2cm} \right),
$$
so that,
\begin{equation}
\bfD_{\bfaX \bfaY} = \left(\hspace{-0.2cm}
\begin{array}{ccccc}
D(\bfx_4,\bfx_1) & D(\bfx_4,\bfx_3) & D(\bfx_4,\bfy_3) & D(\bfx_4,\bfy_4) \\
D(\bfx_5,\bfx_1) & D(\bfx_5,\bfx_3) & D(\bfx_5,\bfy_3) & D(\bfx_5,\bfy_4) \\
D(\bfx_2,\bfx_1) & D(\bfx_2,\bfx_3) & D(\bfx_2,\bfy_3) & D(\bfx_2,\bfy_4) \\
D(\bfx_1,\bfy_2) & D(\bfx_3,\bfy_2) & D(\bfy_2,\bfy_3) & D(\bfy_2,\bfy_4) \\
D(\bfx_2,\bfy_1) & D(\bfx_3,\bfy_1) & D(\bfy_1,\bfy_3) & D(\bfy_1,\bfy_4) \\
\end{array}
\hspace{-0.2cm} \right).
\label{eq:euclid.matrix.xstar.ystar.2}
\end{equation}

Finally, note that because the order of the elements of $\bfax$ in equation \ref{eq:permuted.data.2} is different from the order in equation \ref{eq:permuted.data}, we have that the order of the elements of the $\bfD_{\bfaX \bfaX}$ in equation \ref{eq:euclid.matrix.xstar.xstar.2} is different from \ref{eq:euclid.matrix.xstar.xstar} (and, similarly, the order of the elements $\bfD_{\bfaX \bfaY}$ in equation \ref{eq:euclid.matrix.xstar.ystar.2} is different from equation \ref{eq:euclid.matrix.xstar.ystar}). Different orders, nonetheless, do not affect the value of the ED statistic which only depend on averages over all elements of each of these matrices.


\begin{thebibliography}{50}

\bibitem{Efron1993} Bradley Efron and Robert Tibshirani. 1993. An introduction to the bootstrap. Chapman \& Hall, New York.

\bibitem{Good2000} Phillip I. Good. 2000. Permutation tests: a practical guide to resampling methods for testing hypotheses (2nd ed ed.). Springer, New York.

\bibitem{Gretton2006} Arthur Gretton, Karsten M. Borgwardt, Malte J. Rasch, Bernhard Scholkopf, Alexander J. Smola. 2006. A kernel method for the two-sample-problem. Proceedings of the 20th International Conference on Neural Information Processing Systems, 2006.

\bibitem{Gretton2007} Arthur Gretton, Kenji Fukumizu, Choon Hui Teo, Le Song, Bernhard Scholkopf, Alexander J. Smola. 2007. A kernel statistical test of independence. Proceedings of the 20th International Conference on Neural Information Processing Systems, 2007, 585-592.

\bibitem{Gretton2009} Arthur Gretton, Kenji Fukumizu, Zaid Harchaoui, Bharath K. Sriperumbudur. 2009. A fast, consistent kernel two-sample test. In Advances in Neural Information Processing Systems 22, Red Hook, NY, 2009. Curran Associates Inc

\bibitem{Gretton2012} Arthur Gretton, Karsten M. Borgwardt, Malte J. Rasch, Bernhard Scholkopf, Alexander Smola. 2012. A kernel two-sample test. Journal of Machine Learning Research 13(25):723-773.

\bibitem{rproject2019} R Core Team. 2019. R: A language and environment for statistical computing. R Foundation for Statistical Computing, Vienna, Austria. URL http://www.R-project.org/.

\bibitem{Phipson2010} Belinda Phipson and Gordon K Smyth. 2010. Permutation P-values Should Never Be Zero: Calculating Exact P-values When Permutations Are Randomly Drawn. Stat. Appl. Genet. Mol. Biol. 9, 1 (January 2010). https://doi.org/10.2202/1544-6115.1585

\bibitem{Ramdas2015} Aaditya Ramdas, Barnabas Poczos, Aarti Singh, Larry Wasserman. 2015. Adaptivity and computation-statistics tradeoffs for kernel and distance based high dimensional two sample testing. arXiv:1508.00655

\bibitem{Sejdinovic2013} Dino Sejdinovic, Bharath Sriperumbudur, Arthur Gretton, and Kenji Fukumizu. 2013. Equivalence of distance-based and RKHS-based statistics in hypothesis testing. Ann. Stat. 41, 5 (October 2013). https://doi.org/10.1214/13-AOS1140

\bibitem{SzekelyRizzo2004} Gabor J. Szekely and Maria L. Rizzo. 2004. Testing for Equal Distributions in High Dimension, InterStat, November (5).

\bibitem{SzekelyRizzo2005} Gabor J. Szekely and Maria L. Rizzo. 2005. A new test for multivariate normality. J. Multivar. Anal. 93, 1 (March 2005), 58-80. https://doi.org/10.1016/j.jmva.2003.12.002

\bibitem{SzekelyRizzo2017} Gabor J. Szekely and Maria L. Rizzo. 2017. The Energy of Data. Annu. Rev. Stat. Its Appl. 4, 1 (March 2017), 447-479. https://doi.org/10.1146/annurev-statistics-060116-054026

\bibitem{Shekhar2022} Shubhanshu Shekhar, Ilmun Kim, Aaditya Ramdas. 2022. A permutation-free kernel two-sample test. In Advances in Neural Information Processing Systems 36.

\bibitem{Song2012} Le Song, Alex Smola, Arthur Gretton, Justin Bedo, Karsten Borgwardt. 2012. Feature selection via dependence maximization. Journal of Machine Learning Research. 13(47):1393-1434.

\bibitem{Zaremba2013} Wojciech Zaremba, A. Gretton, and Matthew Blaschko. 2013. B-test: low variance kernel two-sample test. Advances in Neural Information Processing Systems, 2013.

\end{thebibliography}
\end{document}